\documentclass[aps,prd,12pt,%
  nofootinbib,superscriptaddress%
  ,eqsecnum
  ,preprint
]{revtex4}
\usepackage{graphicx}

\renewcommand{\thesection}{\arabic{section}}

\makeatletter
\renewcommand{\p@subsection}{}
\makeatother

\begin{document}

\begin{flushright}
\begin{minipage}{3cm}
\begin{flushleft}
DPNU-03-08
\end{flushleft}
\end{minipage}
\end{flushright}

\title{%
Vector Manifestation and Violation of
   Vector Dominance \\ in Hot Matter
}

\author{Masayasu Harada}
\author{Chihiro Sasaki}
\affiliation{Department of Physics, Nagoya University,
  Nagoya, 464-8602, Japan}

\begin{abstract}

We show the details of the calculation of the hadronic thermal
corrections to the two-point functions in the effective field theory
of QCD for pions and vector mesons based on the hidden local symmetry
(HLS) in hot matter using the background field gauge.  
We study the
temperature dependence of the pion velocity in the low temperature
region determined from the hadronic thermal corrections, and show that,
due to the presence of the dynamical vector meson,
the pion velocity is smaller than the speed of the light already at
one-loop level, in contrast to the result obtained in the ordinary
chiral perturbation theory including only the pion at one-loop.  
Including the intrinsic temperature
dependences of the parameters of the HLS Lagrangian determined from
the underlying QCD through the Wilsonian matching,
we show how the vector manifestation (VM), in which the massless
vector meson becomes the chiral partner of pion, is realized at the
critical temperature.
We present a new prediction of the VM on the
direct photon-$\pi$-$\pi$ coupling which measures the validity of the
vector dominance (VD) of the electromagnetic form factor of the pion:
We find that the VD is largely violated at the critical temperature,
which indicates that the assumption of the VD made in several analyses
on the dilepton spectra in hot matter may need to be weakened for
consistently including the effect of the dropping mass of the vector
meson.

\end{abstract}

\maketitle

\newpage


\section{Introduction}

Spontaneous chiral symmetry breaking is one of the 
most important  
features in low energy QCD.
The phenomena of 
the light pseudoscalar mesons (the pion and its flavor partner), 
which are regarded as
the approximate Nambu-Goldstone bosons  
associated with the symmetry breaking, are well described
by the symmetry property in the low energy region.
This chiral symmetry is expected to be restored in hot and/or
dense matter
and properties of hadrons will be changed near the critical point
of the chiral symmetry restoration
~\cite{HatsudaKunihiro,Pisarski:95,%
Brown-Rho:96,Brown:2001nh,HatsudaShiomiKuwabara,%
Rapp-Wambach:00,Wilczek}.
It is important to investigate the physics 
near the phase transition point both theoretically and experimentally.
In fact, the CERN Super Proton Synchrotron (SPS) observed
an enhancement of dielectron ($e^+e^-$) mass spectra
below the $\rho / \omega$ resonance~\cite{Agakishiev:1995xb}.
This can be explained by the dropping mass of the $\rho$ meson
(see, e.g., Refs.~\cite{Li:1995qm, Brown-Rho:96, Rapp-Wambach:00})
following the Brown-Rho scaling proposed in Ref.~\cite{BR}.
Furthermore,
the Relativistic Heavy Ion Collider (RHIC) has started
to measure several physical processes in hot matter
which include the dilepton energy spectra.
This will further clarify the properties of vector mesons
in hot matter.
Therefore it is interesting to study the properties of the pion and
vector mesons near the critical temperature,
especially associated with the dropping mass of the vector meson.

There is no strong restriction concerning vector meson masses
in the standard scenario of chiral symmetry restoration, where
the pion joins with the scalar meson in the same chiral representation
[see for example, Refs.~\cite{HatsudaKunihiro,Bochkarev:1995gi,PT:96}].
However there is a scenario which certainly requires the dropping mass 
of the vector meson and supports the Brown-Rho scaling:
In Ref.~\cite{HYb}, it was proposed that
there can be another possibility for the pattern 
of chiral symmetry restoration,
named the vector manifestation (VM).
The VM was proposed
as a novel manifestation of the chiral symmetry in the Wigner
realization, in which the chiral symmetry is restored by 
the massless degenerate pion (and its flavor partners) and the longitudinal 
$\rho$ meson (and its flavor partners) as the chiral partner.
In terms of the chiral representations of the low-lying mesons,
there is a representation mixing in the vacuum~\cite{HYb,HYc}.
When we approach the critical point, there are two possibilities
for the pattern of the chiral symmetry restoration:
One possible pattern is the standard scenario and another is the VM.
Both of them are on an equal footing with each other
in terms of the chiral representations.
It is worthwhile to study the physics associated with the VM as well as
that with the standard scenario of the chiral symmetry restoration.

It has been shown that the VM is formulated at a large number of
flavor~\cite{HYb}, critical temperature~\cite{HSasaki} and
critical density~\cite{HKR} by using the effective field theory
including both pions and vector mesons
based on the hidden local symmetry (HLS)~\cite{BKUYY,BKYa},
where a second order or weakly first order phase transition was assumed.
In the VM at finite temperature and/or density, 
the {\it intrinsic temperature and/or density dependences} 
of the parameters of the HLS Lagrangian played important roles
to realize the chiral symmetry restoration consistently:
In the framework of the HLS the equality between 
the axial vector and vector current correlators at the critical point
can be satisfied only if the intrinsic thermal and/or density effects are
included.
The intrinsic temperature and/or density dependences are 
nothing but the information converted from the underlying QCD 
through the Wilsonian matching~\cite{HYa,HSasaki}.
In Ref.~\cite{HKRS},
the predictions of the VM were made on the pion velocity and
the vector and axial vector susceptibilities.
It was stressed that the vector mesons 
in addition to the pseudoscalar mesons become relevant degrees of
freedom near the critical temperature.

In this paper
we shed some light on the validity of the vector dominance (VD)
in hot matter.
In several analyses such as the one on the dilepton spectra 
in hot matter carried out
in Ref.~\cite{Rapp-Wambach:00}, the VD is assumed to be held even in
the high temperature region.
There are several analyses resulting in the dropping mass consistent
with the VD, as shown in Refs.~\cite{GomezNicola:2002tn,Dobado:2002xf}.
On the other hand,
the analysis done in Ref.~\cite{Pisarski} shows that 
the thermal vector meson mass goes up if the VD holds.
Thus, it is interesting to study what the VM predicts on the VD.
In the present analysis we
present a new prediction of the VM in hot matter
on the direct photon-$\pi$-$\pi$ coupling which measures the validity
of the VD of the electromagnetic form factor of the
pion. 
We find that {\it the VM predicts a large violation of the VD at
the critical temperature}.
This indicates that the assumption of 
the VD may need to be weakened, at least in some amounts,
for consistently including the
effect of the dropping mass of the vector meson.

This paper is organized as follows:
In section~\ref{sec:HLS}, 
we show the HLS Lagrangian which we use in the present analysis.
In section~\ref{sec:TPFBFG}, 
we present an entire list of the hadronic thermal 
corrections to the two-point functions 
including the vector and pseudoscalar meson loop contributions 
in the background field gauge.
We think that it is useful to summarize the entire list of the hadronic
thermal corrections
since 
only the relevant part of them was listed in Ref.~\cite{HKRS}.
In section~\ref{sec:ITE}, 
we first give an account of the general idea of the intrinsic thermal
and/or dense effects, which was not explicitly stated before.
Then, following Ref.~\cite{HKRS}, 
we briefly review how to extend the Wilsonian matching
to the version at non-zero temperature 
to incorporate the intrinsic
thermal effect.
In section~\ref{sec:VMM}, 
we show the explicit forms of the pole masses
of the longitudinal and transverse modes of the vector meson
at non-zero temperature.
Including the intrinsic thermal effect through the Wilsonian matching,
we show how the VM is formulated at the critical temperature.
In section~\ref{sec:PDCPV}, we study the temperature dependences of the
temporal and spatial pion decay constants and the pion velocity
in the low temperature region.
The estimation of the value of the critical temperature is also
carried out
in wider range of input parameters than the one used in 
Ref.~\cite{HSasaki}.
In section~\ref{sec:PaVVD}, we study the validity of the VD in hot matter,
and show that the VD is largely violated near critical
temperature.
In section~\ref{sec:SD}, we give a summary and discussions.
Several functions and formulas used in this paper are listed
in Appendices A-D.


\section{\label{sec:HLS}
Hidden Local Symmetry}

In this section, we show the Lagrangian based on the hidden local
symmetry (HLS) which we use in the present analysis.

The HLS model is based on 
the $G_{\rm{global}} \times H_{\rm{local}}$ symmetry,
where $G=SU(N_f)_L \times SU(N_f)_R$ is the chiral symmetry
and $H=SU(N_f)_V$ is the HLS. 
The basic quantities are 
the HLS gauge boson and two matrix valued
variables $\xi_L(x)$ and $\xi_R(x)$
which transform as
 \begin{equation}
  \xi_{L,R}(x) \to \xi^{\prime}_{L,R}(x)
  =h(x)\xi_{L,R}(x)g^{\dagger}_{L,R}\ ,
 \end{equation}
where $h(x)\in H_{\rm{local}}\ \mbox{and}\ g_{L,R}\in
[\mbox{SU}(N_f)_{\rm L,R}]_{\rm{global}}$.
These variables are parameterized as
 \begin{equation}
  \xi_{L,R}(x)=e^{i\sigma (x)/{F_\sigma}}e^{\mp i\pi (x)/{F_\pi}}\ ,
 \end{equation}
where $\pi = \pi^a T_a$ denotes the pseudoscalar Nambu-Goldstone bosons
associated with the spontaneous symmetry breaking of
$G_{\rm{global}}$ chiral symmetry, 
and $\sigma = \sigma^a T_a$ denotes
the Nambu-Goldstone bosons associated with 
the spontaneous breaking of $H_{\rm{local}}$.
This $\sigma$ is absorbed into the HLS gauge 
boson through the Higgs mechanism. 
$F_\pi \ \mbox{and}\ F_\sigma$ are the decay constants
of the associated particles.
The phenomenologically important parameter $a$ is defined as 
 \begin{equation}
  a = \frac{{F_\sigma}^2}{{F_\pi}^2}\ .
 \end{equation}
The covariant derivatives of $\xi_{L,R}$ are given by
\begin{eqnarray}
 D_\mu \xi_L &=& \partial_\mu\xi_L - iV_\mu \xi_L + i\xi_L{\cal{L}}_\mu,
 \nonumber\\
 D_\mu \xi_R &=& \partial_\mu\xi_R - iV_\mu \xi_R + i\xi_R{\cal{R}}_\mu,
\end{eqnarray}
where $V_\mu$ is the gauge field of $H_{\rm{local}}$, and
${\cal{L}}_\mu \ \mbox{and}\ {\cal{R}}_\mu$ are the external
gauge fields introduced by gauging the $G_{\rm{global}}$ symmetry.

The HLS Lagrangian with the lowest derivative terms in the chiral limit
is given by~\cite{BKUYY,BKYa}
 \begin{equation}
  {\cal{L}}_{(2)} = {F_\pi}^2\mbox{tr}\bigl[ \hat{\alpha}_{\perp\mu}
                                      \hat{\alpha}_{\perp}^{\mu}
                                   \bigr] +
       {F_\sigma}^2\mbox{tr}\bigl[ \hat{\alpha}_{\parallel\mu}
                  \hat{\alpha}_{\parallel}^{\mu}
                  \bigr] -
        \frac{1}{2g^2}\mbox{tr}\bigl[ V_{\mu\nu}V^{\mu\nu}
                   \bigr]
\ , \label{eq:L(2)}
 \end{equation}
where $g$ is the HLS gauge coupling,
$V_{\mu\nu}$ is the field strength
of $V_\mu$ and
 \begin{eqnarray}
  \hat{\alpha}_{\perp }^{\mu}
     &=& \frac{1}{2i}\bigl[ D^\mu\xi_R \cdot \xi_R^{\dagger} -
                          D^\mu\xi_L \cdot \xi_L^{\dagger}
                   \bigr] \ ,
\nonumber\\
  \hat{\alpha}_{\parallel}^{\mu}
     &=& \frac{1}{2i}\bigl[ D^\mu\xi_R \cdot \xi_R^{\dagger}+
                          D^\mu\xi_L \cdot \xi_L^{\dagger}
                   \bigr]
\ .
 \end{eqnarray}

In the HLS,
as first pointed 
in Ref.~\cite{Georgi} and
developed further in Refs.~\cite{HYd,Tana,HYa,HYc},
thanks to the gauge symmetry,
a systematic loop expansion can be formally performed with
the vector mesons included in addition to the pseudoscalar mesons.
In this chiral perturbation theory (ChPT) with  HLS the
vector meson mass is considered as small
compared with the chiral symmetry breaking scale 
$\Lambda_\chi$, by assigning ${\cal O}(p)$ to 
the HLS gauge coupling~\cite{Georgi,Tana}: 
\begin{equation}
 g \sim {\cal O}(p).
\end{equation}
According to the entire list shown in Ref.~\cite{Tana},
there are 35 counter terms at ${\cal O}(p^4)$ for general
$N_f$.
However, only three terms are relevant 
when we consider two-point functions
in the chiral limit:
 \begin{equation}
  {\cal{L}}_{(4)} = z_1\mbox{tr}\bigl[ \hat{\cal{V}}_{\mu\nu}
                       \hat{\cal{V}}^{\mu\nu} \bigr] +
                    z_2\mbox{tr}\bigl[ \hat{\cal{A}}_{\mu\nu}
                       \hat{\cal{A}}^{\mu\nu} \bigr] +
                    z_3\mbox{tr}\bigl[ \hat{\cal{V}}_{\mu\nu}
                       V^{\mu\nu} \bigr], \label{eq:L(4)}
 \end{equation}
where
 \begin{eqnarray}
  \hat{\cal{A}}_{\mu\nu}=\frac{1}{2}
                         \bigl[ \xi_R{\cal{R}}_{\mu\nu}\xi_R^{\dagger}-
                                \xi_L{\cal{L}}_{\mu\nu}\xi_L^{\dagger}
                         \bigr]\ ,
  \label{def A mn}\\
  \hat{\cal{V}}_{\mu\nu}=\frac{1}{2}
                         \bigl[ \xi_R{\cal{R}}_{\mu\nu}\xi_R^{\dagger}+
                                \xi_L{\cal{L}}_{\mu\nu}\xi_L^{\dagger}
                         \bigr]\ ,
  \label{def V mn}
 \end{eqnarray}
with ${\cal{R}}_{\mu\nu}\ \mbox{and}\ {\cal{L}}_{\mu\nu}$ being
the field strengths of ${\cal{R}}_{\mu}\ \mbox{and}\ {\cal{L}}_{\mu}$.

In Ref.~\cite{HYa},
based on the ChPT with HLS,
the Wilsonian matching was proposed,
through which the parameters of the HLS Lagrangian are determined 
by the underlying QCD at the matching scale $\Lambda$.
(For a review of the ChPT with HLS, the Wilsonian matching
and VM at zero temperature, see Ref.~\cite{HYc}.)
The Wilsonian matching at $T=0$ with $N_f = 3$,
where the validity of the ChPT with HLS is essential,
was shown to give several predictions in remarkable agreement with 
experiments~\cite{HYa,HYc}.
This strongly suggests that the ChPT with HLS is valid even numerically.
In Ref.~\cite{HSasaki}, 
we applied the ChPT with HLS combined with the Wilsonian matching
to finite temperature.  There the expansion parameter 
is $T/F_\pi(\Lambda)$ instead of $T/F_\pi(0)$ used in the 
standard ChPT.  Since $F_\pi(\Lambda) > F_\pi(0)$, we think that
the present formalism can be applied in the higher
temperature region than the standard ChPT.


\section{Two-Point Functions in Background Field Gauge}
\label{sec:TPFBFG}

In the present approach
hadronic thermal effects are included
by calculating loop contributions of the
pseudoscalar and vector meson.
In Ref.~\cite{HSasaki} we used
hadronic thermal corrections 
to the pion decay constant and vector meson mass
calculated
within the framework of the chiral perturbation theory
with HLS in the Landau gauge.
While in Ref.~\cite{HKRS}
hadronic thermal corrections were
calculated in the background field gauge.
(For the application of the background field gauge to the HLS,
see, e.g., Ref.~\cite{HYc}.)
In this section we show details of the 
calculation of the hadronic thermal corrections 
to the two-point functions 
in the background field gauge.

Let us consider the loop corrections to the
two-point functions at non-zero temperature.
As was done in Ref.~\cite{HKRS} (see also section~\ref{sec:ITE}), 
we neglect the possible
Lorentz symmetry violating effects caused by the intrinsic temperature
dependences of the bare parameters in the present analysis.
So we calculate loop corrections at non-zero temperature from the bare
Lagrangian with Lorentz invariance shown in Eq.~(\ref{eq:L(2)}). 
For this purpose it is convenient to introduce the following
functions:
\begin{eqnarray}
A_0(M^2;T) &\equiv&
T \sum_{n=-\infty}^{\infty}
\int \frac{d^3k}{(2\pi)^3}
\frac{1}{M^2-k^2}
\ ,
\label{def:A0 2}
\\
B_0(p_0,\bar{p};M_1,M_2;T) &\equiv&
T \sum_{n=-\infty}^{\infty}
\int \frac{d^3k}{(2\pi)^3}
\frac{1}{ [M_1^2-k^2] [M_2^2-(k-p)^2] }
\ ,
\label{def:B0 2}
\\
B^{\mu\nu}(p_0,\bar{p};M_1,M_2;T) &\equiv&
T \sum_{n=-\infty}^{\infty}
\int \frac{d^3k}{(2\pi)^3}
\frac{\left(2k-p\right)^\mu \left(2k-p\right)^\nu}{%
 [M_1^2-k^2] [M_2^2-(k-p)^2] }
\ ,
\label{def:Bmunu 2}
\end{eqnarray}
where $\bar{p} = | \vec{p} |$
and the $0$th component of the loop momentum is taken
as $k^0 = i 2 n \pi T$,
while that of the external momentum is taken as 
$p^0 = i 2 n^\prime \pi T$ [$n, n^\prime$: integer].
Using the standard formula (see, e.g., Ref.~\cite{Kap}),
these functions are divided
into two parts as
\begin{eqnarray}
A_0(M;T) &=&
A_{0}^{\rm(vac)}(M)
+ \overline{A}_{0}(M;T) 
\ ,
\label{def:A0 11}
\\
B_{0}(p_0,\bar{p};M_1,M_2;T) &=&
B_{0}^{\rm(vac)}(p;M_1,M_2)
+ \overline{B}_{0}(p_0,\bar{p};M_1,M_2;T)
\ ,
\label{def:B0 11}
\\
B^{\mu\nu}(p_0,\bar{p};M_1,M_2;T) &=&
B^{{\rm(vac)}\mu\nu}(p;M_1,M_2)
+ \overline{B}^{\mu\nu}(p_0,\bar{p};M_1,M_2;T) 
\ ,
\label{def:Bmunu 11}
\end{eqnarray}
where
$A_{0}^{\rm(vac)}$, $B_{0}^{\rm(vac)}$
and $B^{{\rm(vac)}\mu\nu}$
express the quantum corrections given in 
Eqs.~(\ref{A0vac def})--(\ref{Bmnvac def}),
and 
$\overline{A}_{0}$, 
$\overline{B}_{0}$ and
$\overline{B}^{\mu\nu}$ 
the hadronic thermal corrections.
We summarize the explicit forms of the functions
$\overline{A}_{0}$, 
$\overline{B}_{0}$ and
$\overline{B}^{\mu\nu}$ in various limits relevant to the present
analysis in Appendix~\ref{app:LIFT}.
Note that 
the $0$th component of the momentum
$p_0$ in the right-hand-sides 
of the above expressions
is analytically continued to the Minkowski variable:
$p_0$ is understood as $p_0 + i\epsilon$
($\epsilon \rightarrow +0$) for the retarded function and 
$p_0 - i \epsilon$ for the advanced function.
Then,
$A_{0}^{\rm(vac)}$, $B_{0}^{\rm(vac)}$
and $B^{{\rm(vac)}\mu\nu}$ have no explicit temperature dependence,
while they have intrinsic temperature dependence which is introduced
through the Wilsonian matching as we will see later.
In the following
we write the two-point functions of 
$\overline{\cal A}_\mu$-$\overline{\cal A}_\nu$,
$\overline{\cal V}_\mu$-$\overline{\cal V}_\nu$,
$\overline{V}_\mu$-$\overline{V}_\nu$ and
$\overline{V}_\mu$-$\overline{\cal V}_\nu$
as $\Pi_\perp^{\mu\nu}$, $\Pi_\parallel^{\mu\nu}$,
$\Pi_V^{\mu\nu}$ and
$\Pi_{V\parallel}^{\mu\nu}$, respectively.

At non-zero temperature
there are four independent polarization tensors,
which we choose as
defined in Appendix~\ref{app:PTFT}.
We decompose 
the two-point functions
$\Pi_\perp^{\mu\nu}$, $\Pi_V^{\mu\nu}$ 
$\Pi_{V\parallel}^{\mu\nu}$ and $\Pi_\parallel^{\mu\nu}$ 
as
\begin{eqnarray}
 \Pi_\perp^{\mu\nu}
&=&
u^\mu u^\nu \Pi_\perp^t +
   (g^{\mu\nu}-u^\mu u^\nu)\Pi_\perp^s +
   P_L^{\mu\nu}\Pi_\perp^L + P_T^{\mu\nu}\Pi_\perp^T \ ,
\label{Pi perp decomp}
\\
 \Pi_{V}^{\mu\nu}
 &=&
 u^\mu u^\nu \Pi_{V}^t +
    (g^{\mu\nu} - u^\mu u^\nu)\Pi_{V}^s +
    P_L^{\mu\nu}\Pi_{V}^L +
    P_T^{\mu\nu}\Pi_{V}^T
\ , 
\label{eq:RR-decompose}
\\
 \Pi_{V\parallel}^{\mu\nu}
 &=&
 u^\mu u^\nu \Pi_{V\parallel}^t +
    (g^{\mu\nu} - u^\mu u^\nu)\Pi_{V\parallel}^s +
    P_L^{\mu\nu}\Pi_{V\parallel}^L +
    P_T^{\mu\nu}\Pi_{V\parallel}^T
\ , 
\label{eq:RV-decompose}
\\
 \Pi_{\parallel}^{\mu\nu}
 &=& u^\mu u^\nu \Pi_{\parallel}^t +
    (g^{\mu\nu} - u^\mu u^\nu)\Pi_{\parallel}^s +
    P_L^{\mu\nu}\Pi_{\parallel}^L +
    P_T^{\mu\nu}\Pi_{\parallel}^T
\ ,
\label{eq:VV-decompose}
\end{eqnarray}
where $u^\mu=(1,\vec{0})$
and $P_L^{\mu\nu}$ and $P_T^{\mu\nu}$ are defined in
Eq.~(\ref{A.1}).
Similarly, we decompose the function $B^{\mu\nu}$
as
\begin{equation}
 B^{\mu\nu}
 =u^\mu u^\nu B^t +
   (g^{\mu\nu}-u^\mu u^\nu) B^s +
   P_L^{\mu\nu} B^L + P_T^{\mu\nu} B^T
\ .
\label{Bmn decomp}
\end{equation}
Furthermore, similarly to the division of the functions
into one part for expressing the quantum correction
and another for the hadronic thermal correction done in 
Eqs.~(\ref{def:A0 11})--(\ref{def:Bmunu 11}),
we divide the two-point functions into two parts as
 \begin{equation}
    \Pi^{\mu \nu}(p_0,\bar{p};T) =
           \Pi^{\rm{(vac)}\mu \nu}(p_0,\bar{p}) +
  \overline{\Pi}^{\mu \nu}(p_0,\bar{p};T)
 \ .
      \label{eq:TPart}
 \end{equation}
Accordingly each of four components of the two-point functions
defined in Eqs.~(\ref{Pi perp decomp})--(\ref{eq:VV-decompose})
is divided into the one part for
the quantum correction and another for the hadronic thermal
correction.
We note that
all the divergences are included in the zero temperature part
$\Pi^{\rm(vac)}$.

Now, let us present the entire list of the hadronic thermal corrections
to the two-point functions.
(For the explicit forms of the quantum corrections $\Pi^{\rm (vac)}$,
see Appendix~\ref{app:C}.)

\begin{figure}
 \begin{center}
  \includegraphics[width = 12cm]{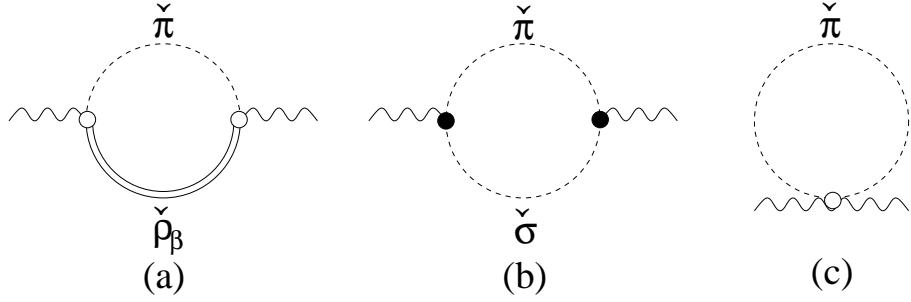}
 \end{center}
 \caption{Diagrams for contributions to $\Pi_\perp^{\mu\nu}$
         at one-loop level. The circle
 $(\circ)$ denotes
         the momentum-independent vertex and the dot
      $(\bullet)$ denotes the momentum-dependent vertex.}
 \label{fig:AAdiagrams}
\end{figure}

We show the diagrams for contributions to
$\Pi_\perp^{\mu\nu}$
at one-loop level in Fig.~\ref{fig:AAdiagrams}.
The temperature dependent parts are obtained as
\begin{eqnarray}
 \overline{\Pi}_\perp^t(p_0,\bar{p};T)
  &=& -N_f a M_\rho^2 \overline{B}_{0}(p_0,\bar{p};M_\rho,0;T) +
     N_f \frac{a}{4}\overline{B}^t(p_0,\bar{p};M_\rho,0;T) \nonumber\\ 
  &&{}+ N_f (a-1) \overline{A}_{0}(0;T),
\label{Pi perp t}
\\
 \overline{\Pi}_\perp^s(p_0,\bar{p};T)
  &=& -N_f a M_\rho^2 \overline{B}_{0}(p_0,\bar{p};M_\rho,0;T) +
     N_f \frac{a}{4}\overline{B}^s(p_0,\bar{p};M_\rho,0;T) \nonumber\\
  &&{}+ N_f (a-1) \overline{A}_{0}(0;T), 
\label{Pi perp s}
\\
 \overline{\Pi}_\perp^L(p_0,\bar{p};T)
   &=& 
  N_f \frac{a}{4}\overline{B}^L(p_0,\bar{p};M_\rho,0;T),
\label{Pi perp L}
\\
 \overline{\Pi}_\perp^T(p_0,\bar{p};T)
   &=& N_f \frac{a}{4}\overline{B}^T(p_0,\bar{p};M_\rho,0;T).
\end{eqnarray}

\begin{figure}
 \begin{center}
  \includegraphics[width = 13cm]{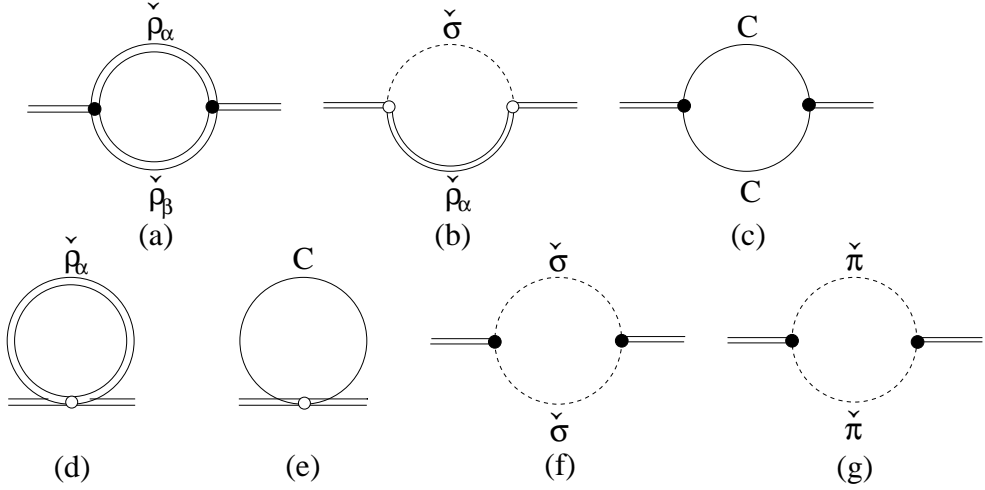}
 \end{center}
 \caption{Diagrams for contributions to $\Pi_V^{\mu\nu}$
         at one-loop level.}
 \label{fig:RRdiagrams}
\end{figure}

As we will see later in section~\ref{sec:VMM},
we define 
the vector meson mass as the pole
of longitudinal or transverse component 
of the vector meson propagator. 
The diagrams for contributions to $\Pi_V^{\mu\nu}$ are shown in 
Fig.~\ref{fig:RRdiagrams}.
We obtain 
the hadronic thermal corrections to 
$\overline{\Pi}_V^{\mu\nu}$
as
\begin{eqnarray}
 \overline{\Pi}_V^t(p_0,\bar{p};T)
 &=& 2N_f\overline{A}_{0}(M_\rho;T) 
    - M_\rho^2 N_f
    \overline{B}_{0}(p_0,\bar{p};M_\rho,M_\rho;T) \nonumber\\
 &&{}+\frac{a^2}{8}N_f
       \overline{B}^t(p_0,\bar{p};0,0;T) 
   {}+\frac{9}{8}N_f
       \overline{B}^t(p_0,\bar{p};M_\rho,M_\rho;T)\ ,
\label{PiRt}
\\
 \overline{\Pi}_V^s(p_0,\bar{p};T)
 &=& 2N_f\overline{A}_{0}(M_\rho;T) 
    - M_\rho^2 N_f
    \overline{B}_{0}(p_0,\bar{p};M_\rho,M_\rho;T) \nonumber\\
 &&{}+\frac{a^2}{8}N_f
       \overline{B}^s(p_0,\bar{p};0,0;T) 
   {}+\frac{9}{8}N_f
       \overline{B}^s(p_0,\bar{p};M_\rho,M_\rho;T)\ ,
\label{PiRs}
\\
 \overline{\Pi}_V^L(p_0,\bar{p};T)
 &=& -4p^2 N_f \overline{B}_{0}(p_0,\bar{p};M_\rho,M_\rho;T)\nonumber\\
 &&{}+\frac{a^2}{8}N_f
       \overline{B}^L(p_0,\bar{p};0,0;T) 
   {}+\frac{9}{8}N_f
       \overline{B}^L(p_0,\bar{p};M_\rho,M_\rho;T),
\label{PiRL}
\\
 \overline{\Pi}_V^T(p_0,\bar{p};T)
 &=& -4p^2 N_f \overline{B}_{0}(p_0,\bar{p};M_\rho,M_\rho;T)\nonumber\\
 &&{}+\frac{a^2}{8}N_f
       \overline{B}^T(p_0,\bar{p};0,0;T) 
   {}+\frac{9}{8}N_f
       \overline{B}^T(p_0,\bar{p};M_\rho,M_\rho;T).
\label{PiRT}
\end{eqnarray}

\begin{figure}
 \begin{center}
  \includegraphics[width = 13cm]{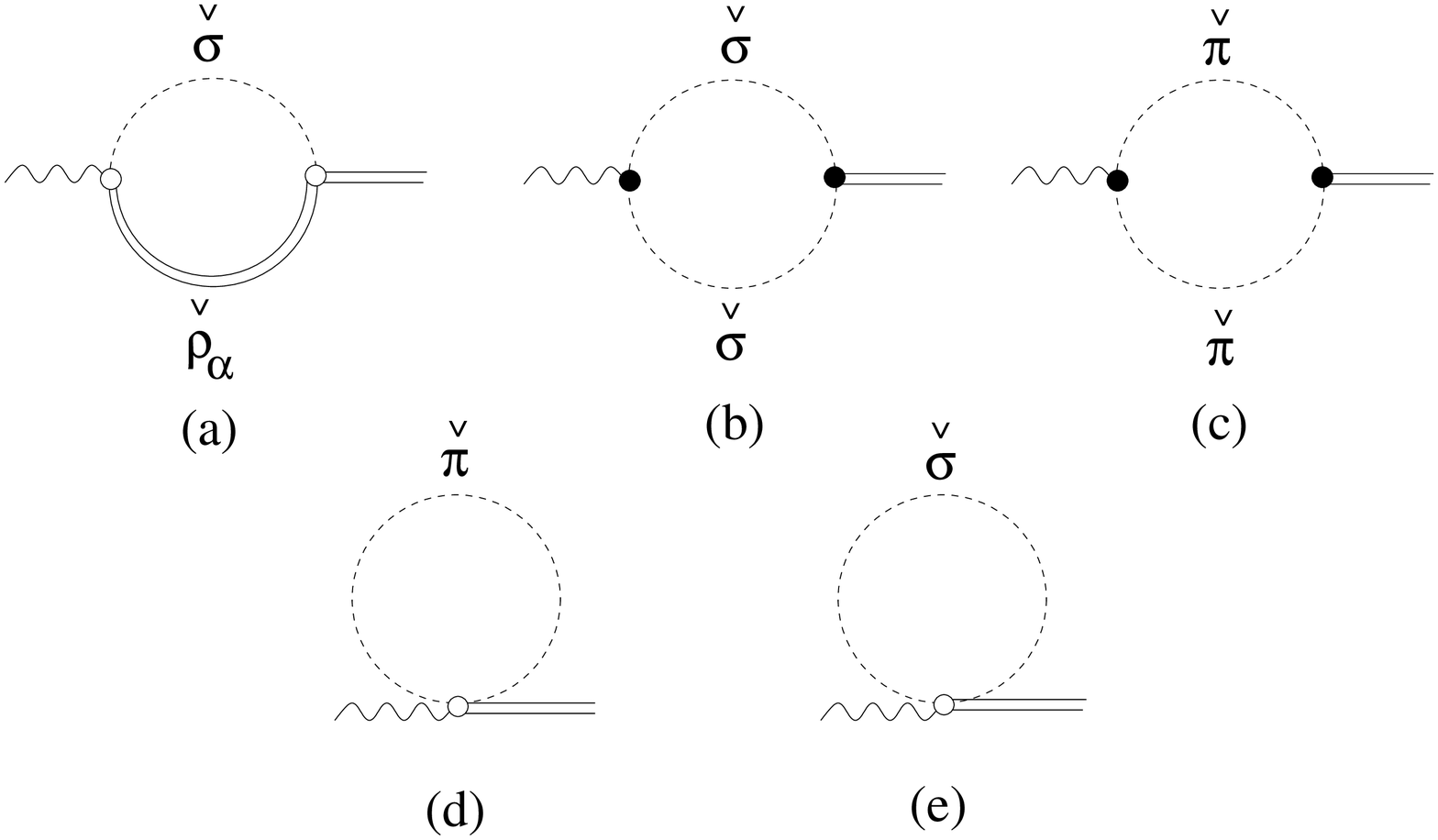}
 \end{center}
 \caption{Diagrams for contributions to
  $\Pi_{V\parallel}^{\mu\nu}$ 
         at one-loop level.}
 \label{fig:VRdiagrams}
\end{figure}

Another two-point function associated with the vector current
correlator 
is $\Pi_{V\parallel}^{\mu\nu}$.
We also show the one-loop diagrams for contributions to
$\Pi_{V\parallel}^{\mu\nu}$ in Fig.~\ref{fig:VRdiagrams}.
We get the temperature dependent parts as
\begin{eqnarray}
 \overline{\Pi}_{V\parallel}^t(p_0,\bar{p};T)
 &=& \frac{a}{2}N_f \overline{A}_{0}(0;T)
   {}+ \frac{1}{2}N_f \overline{A}_{0}(M_\rho;T) 
{}+N_f M_\rho^2 
  \overline{B}_{0}(p_0,\bar{p};M_\rho,M_\rho;T)\nonumber\\
 &&{}+\frac{1}{8}N_f \overline{B}^t(p_0,\bar{p};M_\rho,M_\rho;T)
   {}+\frac{a(2-a)}{8}N_f 
       \overline{B}^t(p_0,\bar{p};0,0;T) ,
\label{PiVt}
\\
 \overline{\Pi}_{V\parallel}^s(p_0,\bar{p};T)
 &=& \frac{a}{2}N_f \overline{A}_{0}(0;T)
   {}+ \frac{1}{2}N_f \overline{A}_{0}(M_\rho;T) 
   {}+N_f M_\rho^2 
      \overline{B}_{0}(p_0,\bar{p};M_\rho,M_\rho;T)\nonumber\\
 &&{}+\frac{1}{8}N_f \overline{B}^s(p_0,\bar{p};M_\rho,M_\rho;T)
   {}+\frac{a(2-a)}{8}N_f \overline{B}^s(p_0,\bar{p};0,0;T),
\label{PiVs}
\\
 \overline{\Pi}_{V\parallel}^L(p_0,\bar{p};T)
 &=& \frac{1}{8}N_f \overline{B}^L(p_0,\bar{p};M_\rho,M_\rho;T)
   {}+\frac{a(2-a)}{8}N_f\overline{B}^L(p_0,\bar{p};0,0;T),
\\
 \overline{\Pi}_{V\parallel}^T(p_0,\bar{p};T)
 &=& \frac{1}{8}N_f \overline{B}^T(p_0,\bar{p};M_\rho,M_\rho;T)
   {}+\frac{a(2-a)}{8}N_f \overline{B}^T(p_0,\bar{p};0,0;T).
\end{eqnarray}

\begin{figure}
 \begin{center}
  \includegraphics[width = 13cm]{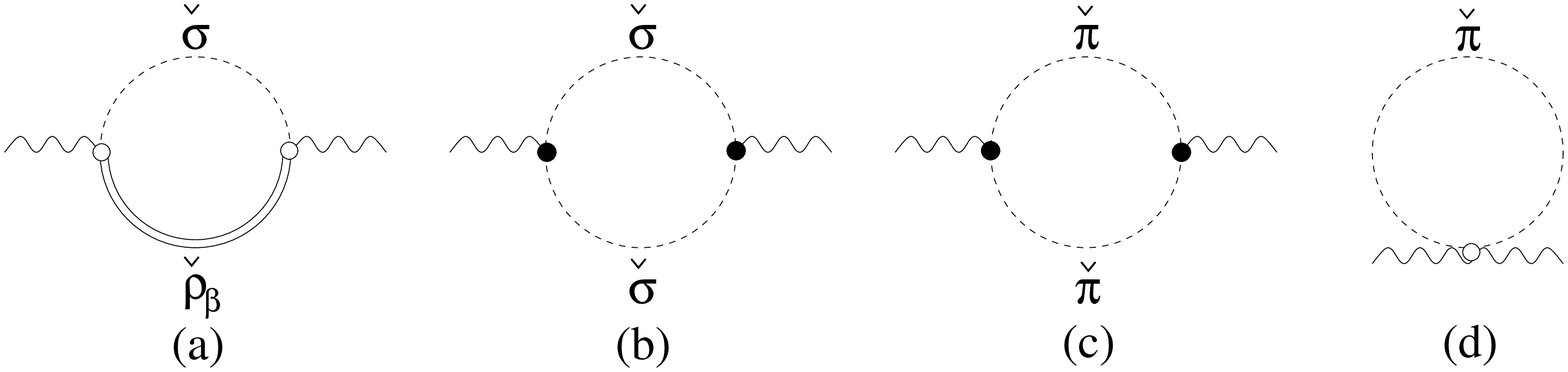}
 \end{center}
 \caption{Diagrams for contributions to
  $\Pi_\parallel^{\mu\nu}$ at one-loop level.}
 \label{fig:VVdiagrams}
\end{figure}

In section~\ref{sec:PaVVD}, we will define the parameter $a$
from the direct $\gamma\pi\pi$ coupling
using the two-point function 
$\Pi_\parallel^{\mu\nu}$.
We show the diagrams for contributions to 
$\Pi_\parallel^{\mu\nu}$
at one-loop level in Fig.~\ref{fig:VVdiagrams}.
We get the temperature dependent parts as
\begin{eqnarray}
 \overline{\Pi}_\parallel^t(p_0,\bar{p};T) 
 &=&-(a-1)N_f \overline{A}_{0}(0;T)
   {}-N_f M_\rho^2 
     \overline{B}_{0}(p_0,\bar{p};M_\rho,M_\rho;T)\nonumber\\
 &&{}+\frac{1}{8}N_f \overline{B}^t(p_0,\bar{p};M_\rho,M_\rho;T)
   {}+\frac{(2-a)^2}{8}N_f \overline{B}^t(p_0,\bar{p};0,0;T) \ , 
\label{PiVRt}
\\
 \overline{\Pi}_\parallel^s(p_0,\bar{p};T) 
 &=&-(a-1)N_f \overline{A}_{0}(0;T)
   {}-N_f M_\rho^2 
    \overline{B}_{0}(p_0,\bar{p};M_\rho,M_\rho;T)\nonumber\\
 &&{}+\frac{1}{8}N_f \overline{B}^s(p_0,\bar{p};M_\rho,M_\rho;T)
   {}+\frac{(2-a)^2}{8}N_f \overline{B}^s(p_0,\bar{p};0,0;T)\ ,
\label{PiVRs}
\\
 \overline{\Pi}_\parallel^L(p_0,\bar{p};T)
 &=& \frac{1}{8}N_f \overline{B}^L(p_0,\bar{p};M_\rho,M_\rho;T)
   {}+\frac{(2-a)^2}{8}N_f\overline{B}^L(p_0,\bar{p};0,0;T),
\\
 \overline{\Pi}_\parallel^T(p_0,\bar{p};T)
 &=& \frac{1}{8}N_f \overline{B}^T(p_0,\bar{p};M_\rho,M_\rho;T)
   {}+\frac{(2-a)^2}{8}N_f \overline{B}^T(p_0,\bar{p};0,0;T).
\end{eqnarray}

At the end of this section, 
using the relations shown in Eq.~(\ref{B.8}),
we obtain
\begin{eqnarray}
&& \overline{\Pi}_V^t =
- \overline{\Pi}_{V\parallel}^t = \overline{\Pi}_{\parallel}^t
\nonumber\\
&& \quad
  = \overline{\Pi}_V^s =
-\overline{\Pi}_{V\parallel}^s = \overline{\Pi}_{\parallel}^s
\nonumber\\
&& \quad
 = - N_f \frac{1}{4} 
  \left[
    \overline{A}_{0}(M_\rho;T) 
     + \overline{A}_{0}(0;T)
  \right]
  - N_f M_\rho^2 \overline{B}_{0}(p_0,\bar{p};M_\rho,M_\rho;T)
\ .
\label{Pi bar V ts}
\end{eqnarray}
Since the quantum corrections to the
corresponding components are identical
as we have shown in Eq.~(\ref{Pi V S equal}),
the above relation implies that 
the components
$\Pi_V^t$, $\Pi_{V\parallel}^t$, $\Pi_\parallel^t$, 
$\Pi_V^s$, $\Pi_{V\parallel}^s$ and $\Pi_\parallel^t$ 
agree:
\begin{equation}
\Pi_V^t = 
- \Pi_{V\parallel}^t = \Pi_{\parallel}^t
  = \Pi_V^s =
-\Pi_{V\parallel}^s = \Pi_{\parallel}^s
\ .
\label{Pi V ts}
\end{equation}
This relation is important to prove the conservation of the
vector current correlator as shown in Ref.~\cite{HKRS}.~\footnote{
  Actually, for the conservation of the vector current correlator,
  $\Pi_V^t = -\Pi_{V\parallel}^t = \Pi_{\parallel}^t$ and
  $\Pi_V^s = -\Pi_{V\parallel}^s = \Pi_{\parallel}^s$ are enough,
  and $\Pi_V^t$ and $\Pi_V^s$ can be generally different.
  }


\section{Intrinsic Thermal Effects}
\label{sec:ITE}

In this section,
we first give an account of the general idea of the intrinsic 
temperature and/or density dependences of the bare parameters.
Next, following Ref.~\cite{HKRS},
we briefly review how to extend the Wilsonian matching
to the version at non-zero temperature in order to incorporate
the intrinsic thermal effect into the bare parameters of the HLS
Lagrangian. 
There we discuss the effect of Lorentz symmetry violation
at bare level, and
summarize the conditions for the bare parameters obtained
in Ref.~\cite{HSasaki} through the Wilsonian matching at the critical
temperature.


\subsection{General Concept of the Intrinsic Thermal and/or
Density Effects}

In this subsection,
we give a general idea of the matching between the effective field
theory (EFT) and QCD.
Green's functions have important information of system,
which is generated by the following functional of a set of 
source fields denoted by $J$:
\begin{equation}
 Z[J] = \int {\cal D}q {\cal D}\bar{q} {\cal D}G
         e^{S_{\rm QCD}[J]},
\label{gf}
\end{equation}
where $q$ $(\bar{q})$ denotes quark or antiquark field,
$G$ is gluon field and $S_{\rm QCD}$ represents the action
expressed in terms of the quarks and gluons.
The basic concept of the EFT is that the effective Lagrangian,
which has the most general form constructed from the chiral symmetry,
can give the same generating functional as in Eq.~(\ref{gf}):
\begin{equation}
 Z[J] = \int {\cal D}U e^{S_{\rm eff}[J]},
\label{gf-eff}
\end{equation}
where $U$ denotes the relevant hadronic fields such as the pion fields
and $S_{\rm eff}$ is the action 
expressed in terms of these hadrons.
Corresponding to each Green's function derived from Eq.~(\ref{gf}),
we have the same Green's function obtained from the EFT
through Eq.~(\ref{gf-eff}).
Performing the matching between these two Green's functions,
we can determine the parameters of the EFT.

In some matching schemes, the renormalized parameters of the EFT
are determined from QCD.
On the other hand,
the matching in the Wilsonian sense is performed based on the 
following general idea:
The bare Lagrangian of the EFT is defined at a suitable 
matching scale $\Lambda$ and
the generating functional derived from the bare Lagrangian 
leads to the same Green's function as that derived from Eq.~(\ref{gf})
at $\Lambda$.
Then the {\it bare} parameters of the EFT are determined 
through the Wilsonian matching.
In other words,
we obtain the bare Lagrangian of the EFT after
integrating out the high energy modes, i.e.,
the quarks and gluons above $\Lambda$.
The information of the high energy modes is included in the parameters
of the EFT.
Thus when we integrate out high energy modes in hot and/or dense matter,
the parameters are in general dependent on temperature and/or density.
The intrinsic temperature and/or density dependences are 
nothing but the signature that the hadron has an internal structure 
constructed from the quarks and gluons.
This is similar to the situation where the coupling constants
among hadrons are replaced with the momentum-dependent form factor
in the high energy region.
Thus the intrinsic thermal and/or dense effects play more important
roles in the higher temperature region, 
especially near the critical temperature.


\subsection{Wilsonian Matching}

The Wilsonian matching proposed in Ref.~\cite{HYa} is done by matching
the axial vector and vector current correlators derived from the
HLS with those from the operator product expansion (OPE) in
QCD at the matching scale $\Lambda$~\footnote{
 For the validity of the expansion in the HLS the
 matching scale $\Lambda$ must be smaller than the chiral symmetry
 breaking scale $\Lambda_\chi$.
}.

Here it should be noticed that
there is no longer Lorentz symmetry
in hot matter,
and
the Lorentz non-scalar operators such as
$\bar{q}\gamma_\mu D_\nu q$ may exist in 
the form of the current correlators derived from the OPE~\cite{HKL}.
This leads to a difference between the temporal and spatial
bare pion decay constants.
However, we neglect the contributions from these operators
since they give a small correction compared with 
the main term $1 + \frac{\alpha_s}{\pi}$~\cite{HSasaki}.
This implies that the Lorentz symmetry breaking effect in
the bare pion decay constant is small, 
$F_{\pi,\rm{bare}}^t \simeq F_{\pi,\rm{bare}}^s$~\cite{HKRS}.
Thus to a good approximation we determine the pion decay
constant at non-zero temperature through the matching 
condition at zero temperature, putting possible 
temperature dependences into the gluonic 
and quark condensates~\cite{HSasaki, HKRS}:
\begin{equation}
 \frac{F^2_\pi (\Lambda ;T)}{{\Lambda}^2} 
  = \frac{1}{8{\pi}^2}\Bigl[ 1 + \frac{\alpha _s}{\pi} +
     \frac{2{\pi}^2}{3}\frac{\langle \frac{\alpha _s}{\pi}
      G_{\mu \nu}G^{\mu \nu} \rangle_T }{{\Lambda}^4} +
     {\pi}^3 \frac{1408}{27}\frac{\alpha _s{\langle \bar{q}q
      \rangle }^2_T}{{\Lambda}^6} \Bigr]
\ .
\label{eq:WMC A}
\end{equation}
Through this condition
the temperature dependences of the quark and gluonic condensates
determine the intrinsic temperature dependences 
of the bare parameter $F_\pi(\Lambda;T)$,
which is then converted into 
those of the on-shell parameter $F_\pi(\mu=0;T)$ 
through the Wilsonian RGEs.

Now, let us consider the Wilsonian matching near the
chiral symmetry restoration point.
Here we assume that the order of the chiral phase transition is
second or weakly first order,
and thus the quark condensate becomes zero
continuously for $T \to T_c$.
First, note that
the Wilsonian matching condition~(\ref{eq:WMC A}) 
provides
\begin{equation}
  \frac{F^2_\pi (\Lambda ;T_c)}{{\Lambda}^2} 
  = \frac{1}{8{\pi}^2}\Bigl[
                            1 + \frac{\alpha _s}{\pi} +
                             \frac{2{\pi}^2}{3}
                            \frac{\langle \frac{\alpha _s}{\pi}
                            G_{\mu \nu}G^{\mu \nu} \rangle_{T_c} }
                             {{\Lambda}^4}
                 \Bigr]
 \neq 0 
\ ,
\label{eq:WMC A Tc}
\end{equation}
which implies that the matching with QCD dictates
\begin{equation}
F^2_\pi (\Lambda ;T_c) \neq 0 
\label{Fp2 Lam Tc}
\end{equation}
even at the critical temperature where the on-shell pion decay constant 
vanishes by adding the quantum corrections through
the RGE including the quadratic divergence~\cite{HYa}
and hadronic thermal corrections,
as we will show in section~\ref{sec:PDCPV}.
As was shown in Ref.~\cite{HKR} for the VM in dense matter,
the Lorentz non-invariant version of
the VM conditions for the bare parameters are obtained 
by the requirement of the equality between the axial vector
and vector current correlators in the HLS,
which should be valid also in hot matter~\cite{HKRS}:
\begin{eqnarray}
 && a_{\rm{bare}}^t \equiv
  \Biggl( \frac{F_{\sigma,\rm{bare}}^t}{F_{\pi,\rm{bare}}^t} \Biggr)^2
  \stackrel{T \to T_c}{\to} 1, \quad
  a_{\rm{bare}}^s \equiv
  \Biggl( \frac{F_{\sigma,\rm{bare}}^s}{F_{\pi,\rm{bare}}^s} \Biggr)^2
  \stackrel{T \to T_c}{\to} 1, \label{EVM a}\\
 && g_{T,\rm{bare}} \stackrel{T \to T_c}{\to} 0, \quad
    g_{L,\rm{bare}} \stackrel{T \to T_c}{\to} 0, \label{EVM g}
\end{eqnarray}
where $a^t_{\rm{bare}}, a^s_{\rm{bare}}, g_{T,\rm{bare}}$ and 
$g_{L,\rm{bare}}$ are the extensions of the parameters
$a_{\rm{bare}}$ and $g_{\rm{bare}}$ in the bare Lagrangian
with the Lorentz symmetry breaking effect included as in Appendix A
of Ref.~\cite{HKR}.

When we use the conditions for the parameters $a^{t,s}$ 
in Eq.~(\ref{EVM a})
and the above result that the Lorentz symmetry violation 
between the bare pion decay constants 
$F_{\pi,\rm{bare}}^{t,s}$ is small, 
we can easily show
that the Lorentz symmetry breaking effect between
the temporal and spatial bare sigma decay constants is also small,
$F_{\sigma,\rm{bare}}^t \simeq F_{\sigma,\rm{bare}}^s$~\cite{HKRS}.
While we cannot determine the ratio $g_{L,\rm{bare}}/g_{T,\rm{bare}}$
through the Wilsonian matching
since the transverse mode of vector meson decouples near
the critical temperature.~\footnote{
 In Ref.~\cite{HKR}, the analysis including the Lorentz
 non-invariance at bare HLS theory was carried out.
 Due to the equality between axial vector and vector current
 correlators, $(g_{L,{\rm bare}}, g_{T,{\rm bare}}) \to (0,0)$ is
 satisfied when we approach the critical point.
 This implies that at the bare level the longitudinal mode becomes the real
 NG boson and couples to the vector current correlator, 
 while the transverse mode decouples.
 Furthermore $g_L \to 0$ is a fixed point for the RGE~\cite{Sasaki:2003qj}.
 Thus in any energy scale the transverse mode decouples from the vector
 current correlator.
}
However this implies that the transverse mode is irrelevant
for the quantities studied in this paper.
Therefore in the present analysis, we set
$g_{L,\rm{bare}}=g_{T,\rm{bare}}$ for simplicity and
use the Lorentz invariant Lagrangian at bare level.
In the low temperature region, the intrinsic temperature dependences
are negligible, so that we also use the 
Lorentz invariant Lagrangian at bare level
as in the analysis by the ordinary chiral
Lagrangian in Ref.~\cite{GL}.

At the critical temperature,
the axial vector and vector current correlators
derived in the OPE
agree with each other for any value of $Q^2$.
Thus we require that
these current correlators in the HLS are
equal at the critical temperature
for any value of $Q^2\ \mbox{around}\ {\Lambda}^2$.
As we discussed above, we start from the Lorentz invariant
bare Lagrangian even in hot matter, and then
the axial vector current correlator $G_A^{\rm{(HLS)}}$ and 
the vector current correlator $G_V^{\rm{(HLS)}}$
are expressed by the same forms
as those at zero temperature with the bare parameters
having the intrinsic temperature dependences:
\begin{eqnarray}
 G^{\rm{(HLS)}}_A (Q^2;T) 
  &=& \frac{F^2_\pi (\Lambda;T)}{Q^2} -
      2z_2(\Lambda;T), \nonumber\\
 G^{\rm{(HLS)}}_V (Q^2;T) 
  &=& \frac{F^2_\sigma (\Lambda;T)[1 - 2g^2(\Lambda;T)z_3(\Lambda;T)]}
           {{M_\rho}^2(\Lambda;T) + Q^2} - 2z_1(\Lambda;T).
  \label{correlator HLS at zero-T}
  \end{eqnarray}
By taking account of the fact 
$F^2_\pi (\Lambda ;T_c) \neq 0$ derived from
the Wilsonian matching condition 
given in Eq.~(\ref{eq:WMC A Tc}),
the requirement 
$G_A^{(\rm{HLS})}=G_V^{(\rm{HLS})}$ is satisfied
only if the following conditions are met~\cite{HSasaki}: 
\begin{eqnarray}
 g(\Lambda ;T) &\stackrel{T \to T_c}{\to}& 0, \label{eq:VMg}\\
 a(\Lambda ;T) = F_\sigma^2(\Lambda;T)/F_\pi^2(\Lambda;T)
               &\stackrel{T \to T_c}{\to}& 1, \label{eq:VMa}\\
 z_1(\Lambda ;T) - z_2(\Lambda ;T)
               &\stackrel{T \to T_c}{\to}& 0. \label{eq:VMz}
\end{eqnarray}
Note that the intrinsic thermal effects act on the parameters 
in such a way that they become the values 
of Eqs.~(\ref{eq:VMg})-(\ref{eq:VMz}).

Through the Wilsonian matching at non-zero temperature mentioned above,
the parameters appearing in the hadronic thermal
corrections calculated in section~\ref{sec:TPFBFG}
have the intrinsic temperature dependences:
$F_\pi$, $a$ and $g$ appearing there should be regarded as
\begin{eqnarray}
F_\pi &\equiv& F_\pi(\mu=0;T) \ ,
\nonumber\\
a &\equiv& a\left(\mu=M_\rho(T);T\right) \ ,
\nonumber\\
g &\equiv& g\left(\mu=M_\rho(T);T\right) \ ,
\end{eqnarray}
where $M_\rho$ is determined from the on-shell condition:
  \begin{equation}
   M_\rho^2 \equiv M_\rho^2(T) = 
   a(\mu=M_\rho ;T)g^2(\mu=M_\rho ;T)F_\pi^2(\mu=M_\rho ;T)\ .
                                  \label{eq:Mdef}
  \end{equation}
{}From the RGEs for $g\ \mbox{and}\ a$ in Eqs.(\ref{eq:RGEg})
and (\ref{eq:RGEa}), 
we find that $(g,a)=(0,1)$ is the fixed point.
Therefore, Eqs.~(\ref{eq:VMg}) and (\ref{eq:VMa}) imply that $g$ and $a$
at the on-shell of the vector meson take the same values:
\begin{eqnarray}
&&
 a\,(\mu = M_\rho(T);T) 
\ \mathop{\longrightarrow}^{T\rightarrow T_c}\  1\ , 
\nonumber\\
&&
 g\,(\mu = M_\rho(T);T) 
\ \mathop{\longrightarrow}^{T\rightarrow T_c}\  0
\ ,
\label{VM a g}
\end{eqnarray}
where the parametric vector meson mass $M_\rho(T)$ is determined from
the condition (\ref{eq:Mdef}).
The above conditions with Eq.~(\ref{eq:Mdef}) imply that
$M_\rho(T)$ also vanishes:
\begin{equation}
M_\rho(T) 
\ \mathop{\longrightarrow}^{T\rightarrow T_c}\  0 \ .
\label{VM Mrho}
\end{equation}


\section{Vector Meson Mass}
\label{sec:VMM}

In Ref.~\cite{HSasaki} we have
shown that the vector manifestation (VM) in hot
matter can actually be formulated 
by using the hadronic thermal correction to the vector meson pole mass
calculated in the
Landau gauge~\cite{HS}.
In Ref.~\cite{HKRS} the background field gauge was used to calculate
the hadronic thermal corrections. 
There the vector meson pole mass was not explicitly studied.
In this section we first define the vector meson pole 
masses of both the longitudinal and transverse modes
at non-zero temperature from the vector current correlator 
in the background field gauge and show
the explicit forms of the hadronic thermal corrections
from the
vector and pseudoscalar meson loop.
Then, including the intrinsic temperature dependences of the
parameters near the critical temperature
determined in the previous section, 
we show that the vector meson mass vanishes at the critical temperature.

Let us define pole masses of longitudinal and transverse modes of
the vector meson 
from the poles of longitudinal and transverse components 
of the vector current correlator in the  rest frame~\cite{HKRS}:
\begin{equation}
 G_V^{\mu\nu}=P_L^{\mu\nu}G_V^L + P_T^{\mu\nu}G_V^T,
\label{eq:G_V}
\end{equation}
where
\begin{eqnarray}
 G_V^L
 &=& \frac{\Pi_V^s \bigl( \Pi_V^L + 2\Pi_{V\parallel}^L \bigr)}
         {\Pi_V^s - \Pi_V^L} + \Pi_\parallel^L, \nonumber\\
 G_V^T
 &=& \frac{\Pi_V^s \bigl( \Pi_V^T + 2\Pi_{V\parallel}^T \bigr)}
         {\Pi_V^s - \Pi_V^T} + \Pi_\parallel^T.
\label{GV LT components}
\end{eqnarray}
Then, the pole masses are obtained as the solutions of the following
on-shell conditions:
\begin{eqnarray}
&&
0 = 
\mbox{Re}
\left[
  \Pi_V^s(p_0=m_{\rho}^L(T),0;T) - 
  \Pi_V^L(p_0=m_{\rho}^L(T),0;T)
\right]
\ ,
\nonumber\\
&&
0 = 
\mbox{Re}
\left[
  \Pi_V^s(p_0=m_{\rho}^T(T),0;T) - 
  \Pi_V^T(p_0=m_{\rho}^T(T),0;T)
\right]
\ ,
\label{rho on shell cond}
\end{eqnarray}
where $m_\rho^L(T)$ and $m_\rho^T(T)$ denote the pole masses of
the longitudinal and transverse modes, respectively.
As we have calculated in section~\ref{sec:TPFBFG},
$\Pi_V^s(p_0,\bar{p};T)$,
$\Pi_V^L(p_0,\bar{p};T)$ and
$\Pi_V^T(p_0,\bar{p};T)$ 
in the HLS at one-loop level
are expressed as
\begin{eqnarray}
\Pi_V^s(p_0,\bar{p};T) 
&=&
F_\sigma^2(M_\rho) + \widetilde{\Pi}_V^S(p^2)
+ \overline{\Pi}_V^{s} (p_0,\bar{p};T)
\ ,
\nonumber\\
\Pi_V^L(p_0,\bar{p};T) 
&=&
\frac{p^2}{g^2(M_\rho)} + \widetilde{\Pi}_V^{LT}(p^2)
+ \overline{\Pi}_V^{L} (p_0,\bar{p};T)
\ ,
\nonumber\\
\Pi_V^T(p_0,\bar{p};T) 
&=&
\frac{p^2}{g^2(M_\rho)} + \widetilde{\Pi}_V^{LT}(p^2)
+ \overline{\Pi}_V^{T} (p_0,\bar{p};T)
\ ,
\label{Pi V s L T forms}
\end{eqnarray}
where the explicit forms of 
the finite renormalization effects
$\widetilde{\Pi}_V^S(p^2)$ and
$\widetilde{\Pi}_V^{LT}(p^2)$ 
are given in Eqs.~(\ref{C.2}) and (\ref{C.3}),
and those of the hadronic thermal effects
$\overline{\Pi}_V^{s} (p_0,\bar{p};T)$,
$\overline{\Pi}_V^{L} (p_0,\bar{p};T)$ and
$\overline{\Pi}_V^{T} (p_0,\bar{p};T)$ are
given in Eqs.~(\ref{Pi bar V ts}), (\ref{PiRL}) and (\ref{PiRT}).
Substituting Eq.~(\ref{Pi V s L T forms}) into
Eq.~(\ref{rho on shell cond}),
we obtain
\begin{eqnarray}
\left[m_\rho^L(T)\right]^2 
&=&
M_\rho^2 
{}+
g^2(M_\rho) \,
\Biggl[
  \mbox{Re}\,\widetilde{\Pi}_V^S( p_0^2 )
  + \mbox{Re}\,\overline{\Pi}_V^{s} (p_0,0;T)
\nonumber\\
&& \qquad\qquad\qquad
  {}- \mbox{Re}\,\widetilde{\Pi}_V^{T}(p_0^2)
  - \mbox{Re}\,\overline{\Pi}_V^{L} (p_0,0;T)
\Biggr]_{p_0 = m_\rho^L(T)}
\ ,
\nonumber\\
\left[m_\rho^T(T)\right]^2 
&=&
M_\rho^2 
{}+
g^2(M_\rho) \,
\Biggl[
  \mbox{Re}\,\widetilde{\Pi}_V^S( p_0^2 )
  + \mbox{Re}\,\overline{\Pi}_V^{s} (p_0,0;T)
\nonumber\\
&& \qquad\qquad\qquad
  {}- \mbox{Re}\,\widetilde{\Pi}_V^{T}(p_0^2)
  - \mbox{Re}\,\overline{\Pi}_V^{T} (p_0,0;T)
\Biggr]_{p_0 = m_{\rho}^T(T)}
\ .
\end{eqnarray}
We can replace $m_\rho^L(T)$ and $m_\rho^T(T)$ with
$M_\rho$ in the hadronic
thermal effects as well as in 
the finite renormalization effect,
since the difference is of higher order.
Then, 
noting that 
$\mbox{Re}\,\widetilde{\Pi}_V^S(p^2=M_\rho^2) =
\mbox{Re}\,\widetilde{\Pi}_V^{LT}(p^2=M_\rho^2) = 0$
as shown in Eq.~(\ref{zero FRE}),
we obtain
\begin{eqnarray}
\left[m_{\rho}^L(T)\right]^2
&=&
M_\rho^2 -
g^2(M_\rho) \, \mbox{Re}
\left[
  \overline{\Pi}_V^L(M_\rho;0;T) 
  - \overline{\Pi}_V^s (M_\rho;0;T)
\right]
\ ,
\label{eq:OnShellL}
\\
\left[m_{\rho}^T(T)\right]^2
&=& M_\rho^2 -
g^2(M_\rho) \, \mbox{Re}
\left[
  \overline{\Pi}_V^T(M_\rho;0;T) 
  - \overline{\Pi}_V^s (M_\rho;0;T)
\right]
\ .
\label{eq:OnSellT}
\end{eqnarray}

Let us compute the explicit form of the pole mass using the
expression of $\overline{\Pi}_V^s$,
$\overline{\Pi}_V^L$ and $\overline{\Pi}_V^T$ calculated in 
section~\ref{sec:TPFBFG}.
Here we note that 
Eqs.~(\ref{BL rest}) and (\ref{BT rest}) imply that 
$\overline{B}^L-\overline{B}^s$ agrees with 
$\overline{B}^T-\overline{B}^s$ in the rest frame. 
Then in the rest frame $\overline{\Pi}_V^L-\overline{\Pi}_V^s$
agrees with $\overline{\Pi}_V^T-\overline{\Pi}_V^s$.
Thus the longitudinal pole mass
is the same as the transverse one:
\begin{equation}
m_\rho^L(T) = m_\rho^T(T)\equiv m_\rho(T) \ .
\end{equation}
By using the low momentum limits of the functions
shown in Eqs.~(\ref{B0B rest}) and (\ref{BL rest}),
$\overline{\Pi}^{L}_V-\overline{\Pi}^{s}_V$ in the rest frame is
expressed as
\begin{eqnarray}
 &&\overline{\Pi}^{L}_V(p_0,\bar{p}=0;T)-
 \overline{\Pi}^{s}_V(p_0,\bar{p}=0;T) \nonumber\\
 &=& N_f 
   \Bigl[ \frac{a^2}{24}\tilde{G}_{2}(p_0;T) - 
           \tilde{J}_{1}^2(M_\rho;T) +
          \bigl( \frac{M_\rho^2}{4}-p_0^2 \bigr)
                  \tilde{F}_{3}^2(p_0;M_\rho;T) +
          \frac{3}{8}\tilde{F}_{3}^4(p_0;M_\rho;T)
   \Bigr], 
\end{eqnarray}
where the functions $\tilde{I}_n$, $\tilde{J}^n_m$, $\tilde{F}^n_m$
and $\tilde{G}^n_m$
are defined in Appendix~\ref{app:Functions}.
Substituting the above expression
into Eq.~(\ref{eq:OnShellL}) and using the relation
\begin{eqnarray}
 -\frac{1}{3M_\rho^2}
 \tilde{F}_{3}^4(M_\rho;M_\rho;T)
 &=& \frac{1}{4} \tilde{F}_{3}^2(M_\rho;M_\rho;T) 
   - \frac{1}{3M_\rho^2} \tilde{J}_{1}^2(M_\rho;T)\ ,
\end{eqnarray}
we find that the vector meson pole mass is expressed as
\begin{eqnarray}
  m^2_\rho (T) 
&=& {M_\rho}^2
    +N_f\,g^2
    \Biggl[- \frac{a^2}{12}\tilde{G}_{2}(M_\rho;T)
      + \frac{4}{5} \tilde{J}^2_{1}(M_\rho;T)
      + \frac{33}{16} M_\rho^2 \tilde{F}^2_{3}(M_\rho;M_\rho;T)
    \Biggr].
\label{eq:Mass} 
 \end{eqnarray}
The contribution in this expression
agrees with the result 
in Ref.~\cite{HSasaki} which is derived from the 
hadronic thermal correction calculated
in the Landau gauge 
in Ref.~\cite{HS} by
taking the on-shell condition (\ref{eq:OnShellL}).~\footnote{%
  The functions used in this paper are related to the ones used
  in Ref.~\cite{HSasaki} 
  as $\tilde{J}_{1}^2 = \frac{1}{2\pi^2}J_1^2$, 
  $\tilde{G}_{2}=\frac{1}{2\pi^2}\bar{G}_2$, and so on.
  }

Before going to the analysis near the critical temperature,
we study the behavior of the pole mass in the low temperature
region, $T \ll M_\rho$.
In this region the functions
$F^n_{m}$ and $J^n_{m}$ are suppressed by $e^{-M_\rho/T}$, 
and thus give negligible contributions.
Since $G_{2} \approx -\frac{{\pi}^4}{15}
                         \frac{T^4}{{M_\rho}^2}$,
the vector meson pole mass increases as $T^4$
in the low temperature region,
dominated by the $\pi$-loop effects:
 \begin{equation}
  {m_\rho}^2(T)\approx {M_\rho}^2 +
                 \frac{N_f {\pi}^2 a}{360 {F_\pi}^2}T^4
  \qquad \mbox{for} \quad T \ll M_\rho.
  \label{m rho small T}
 \end{equation}
Note that the lack of $T^2$-term in the above expression is consistent
with the result by the current algebra 
analysis~\cite{Dey-Eletsky-Ioffe}.

Now, let us study the vector meson pole mass near the critical temperature.
As shown in section~\ref{sec:ITE}, the intrinsic temperature
dependences of the parameters of the HLS Lagrangian determined through
the Wilsonian matching imply that the
parametric vector meson mass $M_\rho$ vanishes at the critical temperature.
Then, near the critical temperature we should take 
$M_\rho \ll T$ 
in Eq.~(\ref{eq:Mass})
instead of $T \ll M_\rho$ which was taken to reach the 
expression in Eq.~(\ref{m rho small T}) in the low temperature region.
Thus, by noting that
\begin{eqnarray}
 \tilde{G}_{2}(M_\rho ;T)
   &\stackrel{M_\rho \to 0}{\to}&\tilde{I}_{2}(T), \nonumber\\
 \tilde{J}^2_{1}(M_\rho ;T)
   &\stackrel{M_\rho \to 0}{\to}&\tilde{I}_{2}(T), \nonumber\\
 {M_\rho}^2\tilde{F}^2_{3}(M_\rho ;M_\rho ;T)
   &\stackrel{M_\rho \to 0}{\to}& 0, 
\end{eqnarray}
the pole mass of the vector meson
at $T \lesssim T_c$ becomes
\begin{eqnarray}
 {m_\rho}^2(T) 
  &=& {M_\rho}^2 + 
      N_f\,g^2 \frac{15-a^2}{12}\tilde{I}_{2}(T) 
\nonumber\\
  &=& {M_\rho}^2 +
      N_f\,g^2 \frac{15 - a^2}{144}T^2.
\label{eq:Mass2}
\end{eqnarray}
In the vicinity of $a \simeq 1$
the hadronic thermal effect gives a positive correction,
and then the vector meson pole mass is actually larger than the
parametric mass $M_\rho$.
However, the intrinsic temperature dependences of the parameters
obtained in section~\ref{sec:ITE} lead to 
$g \to 0$ and $M_\rho \rightarrow0$ for $T \to T_c$.
Then, from  Eq.~(\ref{eq:Mass2}) we conclude that
the pole mass of the vector meson $m_\rho$
also vanishes at the critical temperature:
  \begin{equation}
   m_\rho (T) \to 0 \quad \mbox{for} \ T \rightarrow T_c \ .
  \end{equation}
This implies that the VM is realized 
at the critical temperature,
which is consistent with the picture shown in
Refs.~\cite{BR,Brown-Rho:96,Brown:2001jh,Brown:2001nh}.


\section{\label{sec:PDCPV}
  Pion Parameters
  and the Critical Temperature }

In this section we first study the temperature dependences of
the temporal and spatial pion decay constants and 
pion velocity in the low temperature
region, and briefly review how the two decay constants vanish at the
critical temperature following Ref.~\cite{HKRS}.
We also estimate
the value of the critical temperature 
in a wider range of input parameters than the one used in 
Ref.~\cite{HSasaki}.


\subsection{Pion Decay Constant and Pion Velocity}
\label{ssec:PDC}

Since the Lorentz symmetry is violated at non-zero temperature,
there exist two pion decay constants $f_\pi^t$ and $f_\pi^s$~\cite{PT:96}.
These $f_\pi^t$ and $f_\pi^s$ 
are associated with the temporal and spatial components of 
the axial vector current, respectively.
In the present analysis, they are expressed as~\cite{HKRS}
\begin{eqnarray}
 f_\pi^t(\bar{p};T)
 &=& \frac{\Pi_\perp^t(E,\bar{p};T)}{\tilde{F}(\bar{p};T)}
 = \frac{F_\pi^2 + \overline{\Pi}_\perp^t(\bar{p},\bar{p};T)}
   {\tilde{F}(\bar{p};T)}
\ , 
\nonumber\\
 f_\pi^s(\bar{p};T)
 &=& \frac{\Pi_\perp^s(E,\bar{p};T)}{\tilde{F}(\bar{p};T)}
 = \frac{F_\pi^2 + \overline{\Pi}_\perp^s(\bar{p},\bar{p};T)}
   {\tilde{F}(\bar{p};T)}
\ ,
\label{eq:temporal and spatial}
\end{eqnarray}
where $E$ denotes the pion energy and we replaced $E$ by $\bar{p}$ in
the hadronic thermal corrections $\overline{\Pi}_\perp^t (E,\bar{p};T)$ and
$\overline{\Pi}_\perp^s (E,\bar{p};T)$
since the difference is of higher order.
$\tilde{F}(\bar{p};T)$ is
the wave function renormalization constant of 
the $\pi$ field given by~\cite{MOW,HKRS}
\begin{eqnarray}
 \tilde{F}^2(\bar{p};T)
 &=& \mbox{Re}\Pi_\perp^t(E,\bar{p};T) \nonumber\\
 &=& F_\pi^2 + \mbox{Re}\overline{\Pi}_\perp^t(\bar{p},\bar{p};T) \ .
\label{pi wave renorm}
\end{eqnarray}
By using the above decay constants, the pion velocity at one-loop level
is expressed as~\cite{PT:96,HKRS}
~\footnote{
 This form of the pion velocity looks slightly different from the one
 used in Ref.~\cite{HKRS}.
 However, this is actually equivalent to the one in Ref.~\cite{HKRS}
 at one-loop order, and more convenient to study the temperature
 dependence of the pion velocity in the low temperature region.
}
\begin{equation}
v_\pi^2(\bar{p};T) =
 1 + \frac{\tilde{F}(\bar{p};T)}{F_\pi^2}
  \Bigl[ \mbox{Re}f_\pi^s (\bar{p};T) - \mbox{Re}f_\pi^t (\bar{p};T) 
  \Bigr].
\label{fpts rels}
\end{equation}
In Ref.~\cite{HSasaki}, following Ref.~\cite{Bochkarev:1995gi},
we defined the pion decay constant $f_\pi$ as
the pole residue of the axial vector current correlator.
It is related to the above $f_\pi^t$ and $f_\pi^s$ at one-loop order as
\begin{equation}
f_\pi^2(\bar{p};T) =
  f_\pi^t(\bar{p};T) f_\pi^s(\bar{p};T)
\ .
\label{def:fpi}
\end{equation}
Furthermore we have
\begin{equation}
\widetilde{F}(\bar{p};T) = \mbox{Re} \, f_\pi^t(\bar{p};T)
\ .
\label{pi wave renorm 2}
\end{equation}

Substituting Eqs.~(\ref{A0BM}) and (\ref{A0B0})
into Eqs.~(\ref{Pi perp t}) and (\ref{Pi perp s}),
we obtain 
$\overline{\Pi}_\perp^t$ and $\overline{\Pi}_\perp^s$ for the on-shell pion as
\begin{eqnarray}
 &&\overline{\Pi}^{t}_\perp(p_0 = \bar{p}+i\epsilon,\bar{p};T)
 = N_f(a-1) \tilde{I}_{2}(T) \nonumber\\ 
 &&\qquad{}+ \frac{N_f}{2}a
   \left[ \frac{1}{2}
   \overline{B}^t(\bar{p}+i\epsilon,\bar{p};M_\rho,0;T)
 {}-2M_\rho^2 
   \overline{B}_{0}(\bar{p}+i\epsilon,\bar{p};M_\rho,0;T)
   \right]\ , \label{Pit A on-shell}\\
 &&\overline{\Pi}^{s}_\perp(p_0=\bar{p}+i\epsilon,\bar{p};T)
 = N_f(a-1) \tilde{I}_{2}(T) \nonumber\\ 
 &&\qquad{}+ \frac{N_f}{2}a
  \left[ 
    \frac{1}{2}
      \overline{B}^s(\bar{p}+i\epsilon,\bar{p};M_\rho,0;T)
    {}-2M_\rho^2 
      \overline{B}_{0}(\bar{p}+i\epsilon,\bar{p};M_\rho,0;T)
  \right]\ ,
\label{Pits A on-shell}
\end{eqnarray}
where we put $\epsilon \rightarrow +0$ to make the analytic
continuation for the frequency $p_0=i 2\pi nT$ to the Minkowski
variable.
We show the explicit forms of the functions
$\overline{B}^t$, $\overline{B}^s$ and
$\overline{B}_{0}$ in Eqs.~(\ref{B0 A on-shell}),
(\ref{Bt A on-shell}) and (\ref{Bs A on-shell}).

In general the pion velocity in medium
does not agree with the value at $T=0$
due to the interaction with the heat bath.
Below $T_c$,
since the temporal component does not agree with the spatial one
due to the thermal vector meson effect,
$\overline{\Pi}_\perp^t \neq \overline{\Pi}_\perp^s$,
the pion velocity $v_\pi(\bar{p};T)$ is not the speed of light.
As we will see below, in the framework of HLS the pion velocity receives
a change from the $\rho$-loop effect for $0 < T < T_c$.
When we take the low temperature limit ($T \ll M_\rho$)
and the soft pion limit ($\bar{p} \ll M_\rho$ and $\bar{p} \ll T$),
the real parts of
$\overline{\Pi}_\perp^t$ and $\overline{\Pi}_\perp^s$ are approximated as
\begin{eqnarray}
 \mbox{Re}\overline{\Pi}_\perp^t (p_0 = \bar{p}+i\epsilon,\bar{p};T)
  &\simeq&
   {}-N_f \tilde{I}_{2}(T) + N_f \frac{a}{M_\rho^2}\tilde{I}_{4}(T)
   {}- N_f\,a \sqrt{\frac{M_\rho}{8\pi^3}}\,e^{-M_\rho/T}T^{3/2},
\label{t-low-T}\\
 \mbox{Re}\overline{\Pi}_\perp^s (p_0 = \bar{p}+i\epsilon,\bar{p};T)
  &\simeq&
   {}-N_f \tilde{I}_{2}(T) - N_f \frac{a}{M_\rho^2}\tilde{I}_{4}(T)
   {}+ N_f\,a \sqrt{\frac{M_\rho}{8\pi^3}}\,e^{-M_\rho/T}T^{3/2}. 
\label{s-low-T}
\end{eqnarray}
By using Eqs.~(\ref{t-low-T}) and (\ref{s-low-T}) and neglecting the
terms proportional to the suppression factor $e^{-M_\rho / T}$,
the pion velocity is expressed as
\begin{eqnarray}
 v_\pi^2 (\bar{p};T)
 &\simeq& 1 - N_f \frac{2a}{F_\pi^2 M_\rho^2}\tilde{I}_4 (T)
\nonumber\\
 &=& 1 - \frac{N_f}{15}\,a \pi^2 \frac{T^4}{F_\pi^2 M_\rho^2}
 \, <  1\,.
\end{eqnarray}
This shows that the pion velocity is smaller than the speed of light
already at one-loop level in the HLS due to the $\rho$-loop effect.
This is different from the result obtained in the ordinary
ChPT including only the pion at one-loop [see for example, 
Ref.~\cite{PT:96} and references therein].
Generally, the longitudinal $\rho$ contribution to $\overline{\Pi}_\perp^t$
expressed by $\overline{B}^t$ in Eq.~(\ref{Pit A on-shell}) differs from
that to $\overline{\Pi}_\perp^s$ by $\overline{B}^s$ in Eq.~(\ref{Pits A
on-shell}), which implies that, below the critical temperature,
there always exists a deviation of the pion velocity from
the speed of light due to the longitudinal $\rho$-loop effect:
\begin{equation}
 v_\pi^2 (\bar{p};T) < 1 \qquad \mbox{for} \quad 0 < T < T_c\,.
\label{v_pi<1}
\end{equation}

Next, we study the temporal and spatial pion decay constants in hot matter
defined by Eq.~(\ref{eq:temporal and spatial}).
The imaginary parts of $\overline{\Pi}_\perp^t$ and $\overline{\Pi}_\perp^s$
in the low temperature region
are given by
\begin{eqnarray}
 \mbox{Im}\overline{\Pi}_\perp^t (p_0 = \bar{p}+i\epsilon,\bar{p};M_\rho,0;T)
 &\stackrel{\bar{p}\ll T}{\simeq}&
  \frac{N_f}{4}\,a\,\mbox{Im}\overline{B}^t (p_0 = \bar{p}+i\epsilon,\bar{p};
   M_\rho,0;T) \nonumber\\
 &\stackrel{T \ll M_\rho}{\simeq}&
  \frac{N_f}{8\pi}\,a\,M_\rho^2 e^{-M_\rho/T}, \\
 \mbox{Im}\overline{\Pi}_\perp^s (p_0 = \bar{p}+i\epsilon,\bar{p};M_\rho,0;T)
 &\stackrel{\bar{p}\ll T}{\simeq}&
  \frac{N_f}{4}\,a\,\mbox{Im}\overline{B}^s (p_0 = \bar{p}+i\epsilon,\bar{p};
   M_\rho,0;T) \nonumber\\
 &\stackrel{T \ll M_\rho}{\simeq}&
  {}- \frac{N_f}{8\pi}\,a\,M_\rho^2 e^{-M_\rho/T}. 
\end{eqnarray}
Thus the contributions from the imaginary parts 
$\mbox{Im}\overline{\Pi}_\perp^{t,s}$ are small because of the suppression
factor $e^{-M_\rho/T}$.
From Eqs.~(\ref{t-low-T}) and (\ref{s-low-T}) with
$\tilde{I}_{2}(T)=T^2/12$ and $\tilde{I}_{4}(T)=\pi^2T^4/30$,
we obtain the following results for the real parts of 
$f_\pi^t$ and $f_\pi^s$:
\begin{eqnarray}
 \bigl[ \mbox{Re}f_\pi^t \bigr] \tilde{F}
 &\simeq& F_\pi^2 - N_f \Biggl[ \,\frac{T^2}{12} - 
       \frac{a\,\pi^2}{30 M_\rho^2}T^4 + 
       a \sqrt{\frac{M_\rho}{8\pi^3}}\,e^{-M_\rho/T}T^{3/2} \,\Biggr],
\nonumber\\
 \bigl[ \mbox{Re}f_\pi^s \bigr] \tilde{F}
 &\simeq& F_\pi^2 - N_f \Biggl[ \,\frac{T^2}{12} + 
       \frac{a\,\pi^2}{30 M_\rho^2}T^4 -
       a \sqrt{\frac{M_\rho}{8\pi^3}}\,e^{-M_\rho/T}T^{3/2} \,\Biggr].
\label{fpi-ts low-T}
\end{eqnarray}
We note that $\tilde{F}=\mbox{Re}f_\pi^t$ as shown in 
Eq.~(\ref{pi wave renorm 2}).
Then, neglecting the terms suppressed by $e^{-M_\rho / T}$, 
we obtain the difference between $(f_\pi^t)^2$ and $f_\pi^t f_\pi^s$ as
\begin{equation}
 (f_\pi^t)^2 - f_\pi^t f_\pi^s
 \simeq \frac{N_f}{15}\,a \pi^2 \frac{T^4}{M_\rho^2} > 0.
\label{fpit fpits diff}
\end{equation}
This implies that the contribution from the $\rho$-loop 
(the second and third terms in the brackets)
generates a difference between the temporal and spatial pion decay constants
in the low temperature region.
Similarly, at general temperature below $T_c$,
there exists a difference between $f_\pi^t \tilde{F}$ and $f_\pi^s \tilde{F}$
due to the $\rho$-loop effects: 
$\bigl[\mbox{Re}f_\pi^t \bigr] \tilde{F} > 
 \bigl[\mbox{Re}f_\pi^s \bigr] \tilde{F}$.
Since we can always take $\tilde{F}$ to be positive,
we find that $\mbox{Re}f_\pi^t$ is larger than 
$\mbox{Re}f_\pi^s$:
\begin{equation}
 \mbox{Re}f_\pi^t (\bar{p};T) > \mbox{Re}f_\pi^s (\bar{p};T) 
 \qquad \mbox{for} \quad 0 < T < T_c\,.
\label{t>s}
\end{equation}
This result is consistent with Eq.~(\ref{v_pi<1})
because $v_\pi^2 - 1 = (\tilde{F}/F_\pi^2)
[\mbox{Re}f_\pi^s - \mbox{Re}f_\pi^t]$ by definition.
The difference between 
$(f_\pi^t)^2$ and $f_\pi^t f_\pi^s$ shown in 
Eq.~(\ref{fpit fpits diff})
is tiny, and the hadronic thermal corrections to them are
dominated by the first term ($T^2/12$) in the bracket 
in Eq.~(\ref{fpi-ts low-T}).
Then, in the very low temperature region,
the above expressions are further reduced to
\begin{equation}
 f_\pi^2 
  = \bigl( f_\pi^t \bigr)\bigl( f_\pi^s \bigr)
  \sim \bigl( f_\pi^t \bigr)^2
   \sim  {F_\pi}^2 - \frac{N_f}{12}T^2, 
                 \label{eq:f_piGL}
\end{equation}
which is consistent with the result obtained in Ref.~\cite{GL}.

Now we study how $f_\pi^t$ and $f_\pi^s$ behave
near $T_c$.
As we have shown in Eq.~(\ref{eq:WMC A Tc}),
the Wilsonian matching determines the bare pion decay constant
in terms of the parameters appearing in the OPE.
Furthermore, as we discussed around Eq.~(\ref{VM a g}),
at the critical temperature the intrinsic thermal effects lead to
$(g,a) \to (0,1)$ which is a fixed point of the coupled RGEs.
Then the RGE for $F_\pi$ becomes
        \begin{equation}
         \mu \frac{d{F_\pi}^2}{d\mu} = \frac{N_f}{(4\pi)^2}{\mu}^2.
        \end{equation}
This RGE is easily solved and
the relation between $F_\pi (\Lambda ;T_c)\ \mbox{and}\ 
F_\pi (0;T_c)$ is given by
        \begin{equation}
         F^2_\pi (0;T_c) = F^2_\pi (\Lambda ;T_c) -
                           \frac{N_f}{2(4\pi)^2}{\Lambda}^2.
              \label{eq:F_pi0}
        \end{equation}
Finally we obtain the pion decay constants as follows:
\begin{eqnarray}
 f_\pi^t \tilde{F} = F^2_\pi (0;T) + 
        \overline{\Pi}_\perp^t(\bar{p},\bar{p};T), \nonumber\\
 f_\pi^s \tilde{F} = F^2_\pi (0;T) + 
        \overline{\Pi}_\perp^s(\bar{p},\bar{p};T), \label{eq:f_pi}
\end{eqnarray}
where the second terms are the hadronic thermal effects.
In the VM limit $(g,a) \to (0,1)$ the temperature dependent parts become
\begin{equation}
  \overline{\Pi}_\perp^t(\bar{p},\bar{p};T) \stackrel{T \to T_c}{\to}
  {} - \frac{N_f}{24}{T_c}^2, \quad
  \overline{\Pi}_\perp^s(\bar{p},\bar{p};T) \stackrel{T \to T_c}{\to}
  {} - \frac{N_f}{24}{T_c}^2. \label{eq:hadronicT_c}
\end{equation}
{}From Eqs.~(\ref{def:fpi}), (\ref{eq:f_pi}) and (\ref{eq:hadronicT_c}), 
the order parameter $f_\pi$ becomes
\begin{equation}
 f_\pi^2(\bar{p};T) \stackrel{T \to T_c}{\to}
 f_\pi^2(\bar{p};T_c) = 
 F_\pi^2(0;T_c) - \frac{N_f}{24}T_c^2.
\label{fpi critical}
\end{equation}
Since the order parameter $f_\pi$ vanishes at the critical temperature,
this implies
\begin{equation}
 F_\pi^2(0;T) \stackrel{T \to T_c}{\to}
  F_\pi^2(0;T_c) = \frac{N_f}{24}T_c^2. \label{eq:intrinsic}
\end{equation}
Thus we obtain
\begin{equation}
 (f_\pi^t)^2 \stackrel{T \to T_c}{\to} 0, \quad
 f_\pi^t f_\pi^s \stackrel{T \to T_c}{\to} 0.
\end{equation}
{}From Eq.~(\ref{t>s}) or equivalently Eq.~(\ref{v_pi<1}),
the above results imply that the temporal and spatial 
pion decay constants vanish simultaneously 
at the critical temperature~\cite{HKRS}:
\begin{equation}
 f_\pi^t(T_c) = f_\pi^s(T_c) = 0.
\end{equation}
Comparing Eq.~(\ref{fpi critical}) with the expression 
in the low temperature region in Eq.~(\ref{eq:f_piGL})
where the vector meson is decoupled,
we find that the coefficient of ${T_c}^2$ is different
by the factor $\frac{1}{2}$.
This is the contribution from the $\sigma$-loop 
(longitudinal $\rho$-loop).
In the low temperature region
the $\pi$-loop effects give the dominant contributions
to $f_\pi (T)$ and the $\rho$-loop effects are negligible.
However by the vector manifestation
the $\rho$-loop contributions are also incorporated
into $f_\pi (T)$ near the critical temperature.


\subsection{Critical Temperature}
In this subsection we estimate the value of the critical temperature
$T_c$
where the order parameter $f_\pi^2$ vanishes.

We first determine $T_c$
by naively extending
the expression (\ref{eq:f_piGL}) to the higher temperature region
to get $ T_c^{(\rm{hadron})} = 180\,\mbox{MeV}$ for $N_f = 3$.
However this naive extension is inconsistent with 
the chiral restoration in QCD
since the axial vector and vector current correlators do not agree with
each other at that temperature.
As is stressed in Ref.~\cite{HSasaki},
the disagreement between two correlators is cured by including
the intrinsic thermal effect.
As can be seen from Eq.~(\ref{eq:WMC A}),
the intrinsic temperature dependence of the parameter $F_\pi$
is determined from  
${\langle \frac{\alpha _s}{\pi}G_{\mu \nu}G^{\mu \nu}
\rangle}_T$ and  ${\langle \bar{q}q \rangle}_T $,
and gives only a small contribution compared with the main term
$1 + \frac{\alpha_s}{\pi}$.
However it is important that 
the parameters in the hadronic corrections
have the intrinsic temperature dependences as 
$(a,g) \to (1,0)$ for $T \to T_c$,
which carry the information of QCD. 
Actually the inclusion of the intrinsic thermal effects provides
the formula (\ref{fpi critical}) for
the pion decay constant at the critical temperature,
in which the second term has an extra factor of $1/2$ compared with
the one in Eq.~(\ref{eq:f_piGL}).

In Ref.~\cite{HSasaki} we determined the critical temperature
from $f_\pi (T_c) = 0$ and estimated the value which is dependent on
the matching scale $\Lambda$:
{}From Eq.~(\ref{eq:intrinsic}) 
we obtain
  \begin{equation}
   T_c = \sqrt{\frac{24}{N_f}}F_\pi (0;T_c).
  \end{equation}
Using Eqs.(\ref{eq:WMC A Tc}) and (\ref{eq:F_pi0}) we get~\cite{HSasaki}
  \begin{equation}
   \frac{T_c}{\Lambda} = \sqrt{\frac{3}{N_f {\pi}^2}}
         \Bigl[ 1 + \frac{\alpha _s}{\pi} +
                \frac{2{\pi}^2}{3}
                   \frac{\langle \frac{\alpha _s}{\pi}
                     G_{\mu \nu}G^{\mu \nu} \rangle_{T_c}}{{\Lambda}^4} -
         \frac{N_f}{4} \Bigr]^{\frac{1}{2}}. \label{eq:T_c}
  \end{equation}
We would like to stress that the critical temperature is expressed
in terms of the parameters appearing in the OPE.
We evaluate the critical temperature for $N_f=3$.
The gluonic condensate at $T_c$ is about half of the value
at $T=0$~\cite{Miller,Brown:2001nh}
and we use 
$\langle \frac{\alpha_s}{\pi}G_{\mu\nu}G^{\mu\nu} \rangle_{T_c}
= 0.006\,\mbox{GeV}^4$.
We show the predicted values of $T_c$ for several choices of
$\Lambda_{\rm{QCD}}$ and $\Lambda$ in Table \ref{fig:T_c}.
\begin{table}
 \begin{center}
  \begin{tabular}{|c||c|c|c|c|c|c|c|c|c|c|c|c|c|c|c|c|}
    \hline
    $\Lambda_{\rm{QCD}}$ & \multicolumn{3}{c}{} & 0.30 &
                           \multicolumn{3}{c}{} & 0.35 &
                           \multicolumn{3}{c}{} & 0.40 &
                           \multicolumn{3}{c}{} & 0.45 \\
    \hline
    $\Lambda$ & 0.8 & 0.9 & 1.0 & 1.1 &
                0.8 & 0.9 & 1.0 & 1.1 &
                0.8 & 0.9 & 1.0 & 1.1 &
                0.8 & 0.9 & 1.0 & 1.1 \\
    \hline
    $T_c$ & 0.20 & 0.20 & 0.22 & 0.23 &
            0.20 & 0.21 & 0.22 & 0.24 &
            0.21 & 0.22 & 0.23 & 0.25 &
            0.22 & 0.23 & 0.24 & 0.25 \\
    \hline
  \end{tabular}
 \end{center}
 \caption[The critical temperature]
          {Values of the critical temperature for several choices of
           $\Lambda_{\rm{QCD}}\ \mbox{and}\ \Lambda$.
           The units of $\Lambda_{\rm{QCD}}, \Lambda \ \mbox{and}\ 
           T_c$ are GeV.}
 \label{fig:T_c}
\end{table}
Note that the Wilsonian matching describes
the experimental results very well
for $\Lambda_{QCD}=0.4\,\mbox{GeV}$ and $\Lambda=1.1\,\mbox{GeV}$
at $T=0$~\cite{HYc}.
At non-zero temperature, however,
the matching scale $\Lambda$ may be dependent 
on temperature.
The smallest $\Lambda$ in Table~{\ref{fig:T_c}} is determined 
by requiring 
$(2\pi^2/3) \,\langle \frac{\alpha_s}{\pi}G_{\mu\nu}G^{\mu\nu} 
\rangle/ \Lambda^4\,< 0.1$.
Here we note that the critical temperature may be changed by
including the higher order corrections.


\section{Parameter $a$ and Violation of Vector Dominance}
\label{sec:PaVVD}

In this section we study the validity 
of vector dominance (VD) of electromagnetic form factor of the pion
in hot matter.
In Ref.~\cite{Harada:2001rf} 
it has been shown that VD is 
accidentally satisfied in $N_f=3$ QCD at zero temperature and zero
density, and that it is largely 
violated in large $N_f$ QCD when the VM occurs.
At non-zero temperature there exists the hadronic thermal correction
to the parameters.
Thus it is nontrivial whether or not the
VD is realized in hot matter,
especially near the critical temperature.
Here we will show that the intrinsic temperature dependences
of the parameters
of the HLS Lagrangian play essential roles, and then
the VD is largely violated near the critical temperature.

We first study the direct $\gamma\pi\pi$ interaction at zero
temperature. 
We expand the Lagrangian~(\ref{eq:L(2)}) in terms of 
the $\pi$ field with taking the unitary gauge of the 
HLS ($\sigma = 0$) to obtain
\begin{eqnarray}
{\cal{L}}_{(2)} 
&=&
\mbox{tr} \left[ \partial_\mu \pi \partial^\mu \pi \right]
+
a g^2 F_\pi^2 \, \mbox{tr} \left[ \rho_\mu \rho^\mu \right]
+ 2 i \left( \frac{1}{2} a g \right) \, \mbox{tr} 
\left[ \rho^\mu \left[ \partial_\mu \pi \,,\, \pi \right] \right]
\nonumber\\
&&
{}- 2 \left( a g F_\pi^2 \right)
\,\mbox{tr} \left[ \rho_\mu {\cal V}^\mu \right] 
{}+ 2i \left( 1 - \frac{a}{2} \right) \mbox{tr} 
\left[ 
  {\cal V}^\mu \left[ \partial_\mu \pi \,,\, \pi \right] 
\right]
+ \cdots
\ ,
\label{Lag:expand}
\end{eqnarray}
where
the vector meson field $\rho_\mu$ is introduced by
\begin{equation}
V_\mu = g \rho_\mu \ ,
\end{equation}
and 
vector external gauge field ${\cal V}_\mu$ is defined as
\begin{equation}
{\cal V}_\mu \equiv \frac{1}{2} \left(
  {\cal R}_\mu + {\cal L}_\mu
\right)
\ .
\end{equation}
At the leading order of the derivative expansion in the HLS,
the form of the direct $\gamma\pi\pi$ interaction is easily read from
Eq.~(\ref{Lag:expand}) as
\begin{equation}
 \Gamma_{\gamma\pi\pi{\rm(tree)}}^\mu 
  = e(q - k)^\mu(1-\frac{a}{2})\ ,
            \label{eq:gpp}
\end{equation}
where 
$e$ is the electromagnetic coupling constant and
$q$ and $k$ denote outgoing momenta of the pions. 
This shows that,
for the parameter choice $a=2$,
the direct $\gamma\pi\pi$ coupling vanishes
(the VD of the electromagnetic form factor of the pion).

At the next order there exist quantum corrections.
In the background field gauge adopted in the present analysis the
background fields $\overline{\cal A}_\mu$ and 
$\overline{\cal V}_\mu$ include the 
photon field $A_\mu$ and the background pion field $\bar{\pi}$ as
\begin{eqnarray}
\overline{\cal A}_\mu &=&
  \frac{1}{F_\pi(0)}\partial_\mu \bar{\pi} 
  + \frac{i\,e}{F_\pi(0)} A_\mu \left[ Q\,,\, \bar{\pi} \right]
  + \cdots \ ,
\nonumber\\
\overline{\cal V}_\mu &=&
  e \, Q A_\mu 
  - \frac{i}{2F_\pi^2(0)}
    \left[ \partial_\mu \bar{\pi} \,,\, \bar{\pi} \right]
  + \cdots \ ,
\label{expand V A}
\end{eqnarray}
where 
$F_\pi(0)$ is the on-shell pion decay constant
(residue of the pion pole) in order to identify the field $\bar{\pi}$ with the
on-shell pion field, and
the charge matrix $Q$ is given by
\begin{equation}
Q = 
\left(\begin{array}{ccc}
2/3 &      &  \\
    & -1/3 &  \\
    &      & -1/3
\end{array}\right)
\ .
\label{charge}
\end{equation}
The direct $\gamma\pi\pi$ interaction including 
the next order correction is determined from the 
two-point functions of
$\overline{\cal V}_\mu$-$\overline{\cal V}_\nu$ and 
$\overline{\cal A}_\mu$-$\overline{\cal A}_\nu$
and three-point function of 
$\overline{\cal V}_\mu$-$\overline{\cal A}_\alpha$-$%
\overline{\cal A}_\beta$.
We can easily show that contribution from the
three-point function vanishes in the low energy limit as follows:
Let $\Gamma_{\overline{\cal V}\overline{\cal A}\overline{\cal A}}
  ^{\mu\alpha\beta}$
denote the 
$\overline{\cal V}_\mu$-$\overline{\cal A}_\alpha$-$%
\overline{\cal A}_\beta$
three-point function.
Then the direct $\gamma\pi\pi$ coupling 
derived from this three-point function
is proportional to 
$q_{\alpha} k_{\beta} 
 \Gamma_{\overline{\cal V}\overline{\cal A}\overline{\cal A}}
  ^{\mu\alpha\beta}$.
Since the legs $\alpha$ and $\beta$ of
$\Gamma_{\overline{\cal V}\overline{\cal A}\overline{\cal A}}
  ^{\mu\alpha\beta}$ are carried by $q$ or $k$, 
$q_{\alpha} k_{\beta} 
 \Gamma_{\overline{\cal V}\overline{\cal A}\overline{\cal A}}
  ^{\mu\alpha\beta}$ is generally proportional to two of
$q^2$, $k^2$ and $q\cdot k$.
Since the loop integral does not generate any massless poles,
this implies that
$q_{\alpha} k_{\beta} 
 \Gamma_{\overline{\cal V}\overline{\cal A}\overline{\cal A}}
  ^{\mu\alpha\beta}$
vanishes in the low energy limit
$q^2 = k^2 = q \cdot k = 0$.
Thus, the direct $\gamma\pi\pi$ interaction in the low energy limit
can be read from the two-point functions of
$\overline{\cal V}_\mu$-$\overline{\cal V}_\nu$ and 
$\overline{\cal A}_\mu$-$\overline{\cal A}_\nu$ as
\begin{equation}
\Gamma_{\gamma\pi\pi}^\mu
=  e \frac{1}{F_\pi^2(0)}
  \left[ 
    q_\nu \Pi_\perp^{\mu\nu}(q) - k_\nu \Pi_\perp^{\mu\nu}(k)
    - \frac{1}{2} (q - k)_\nu \Pi_\parallel^{\mu\nu}(p)
  \right]
\ ,
\label{gpp T0 0}
\end{equation}
where $p_\nu = (q+k)_\nu$ is the photon momentum.
Substituting the decomposition of the two-point function given in
Eq.~(\ref{decomp T0}) and taking the low-energy limit
$q^2=k^2=p^2=0$, we obtain
\begin{equation}
\Gamma_{\gamma\pi\pi}^\mu
=  e (q-k)^\mu 
  \left[ 
    1
    - \frac{1}{2} \frac{\Pi_\parallel^{{\rm (vac)}S}(p^2=0)}{F_\pi^2(0)}
  \right]
\ ,
\label{gpp T0}
\end{equation}
where we used $\Pi_\perp^{{\rm (vac)}S}(q^2=0) = F_\pi^2(0)$.
Comparing the above expression with the one in Eq.~(\ref{eq:gpp}),
we define the parameter $a(0)$ at one-loop level as
\begin{equation}
 a(0) = \frac{\Pi_\parallel^{{\rm (vac)}S} (p^2=0)}
          {F_\pi^2(0)}\ . \label{eq:Defa}
\end{equation}
We note that, in Ref.~\cite{HYb},
$a(0)$ is defined by the ratio $F_\sigma^2(M_\rho)/F_\pi^2(0)$
by neglecting the finite renormalization effect
which depends on the details of the renormalization condition.
While
the above $a(0)$ in Eq.~(\ref{eq:Defa}) 
defined from the direct $\gamma\pi\pi$ interaction
is equivalent to the one 
used in Ref.~\cite{HKRS}.
In the present renormalization condition (\ref{Fs ren cond}),
$\Pi_\parallel^{{\rm (vac)}S}(p^2=0)$
is given by
\begin{equation}
\Pi_\parallel^{{\rm (vac)}S} (p^2=0) 
=
F_\sigma^2(M_\rho) +
\frac{N_f}{(4\pi)^2} M_\rho^2
\bigl( 2 - \sqrt{3}\tan^{-1}\sqrt{3} \bigr)
\ .
\end{equation}
By adding this, the parameter $a(0)$ is expressed as
\begin{equation}
  a(0) =
  \frac{F_\sigma^2(M_\rho)}{ F_\pi^2(0) }
  {}+\frac{N_f}{(4\pi)^2}\frac{M_\rho^2}{F_\pi^2(0)}
     \bigl( 2 - \sqrt{3}\tan^{-1}\sqrt{3} \bigr) 
\ .
\label{a0 expression}
\end{equation}
Using $M_\rho=771.1\,\mbox{MeV}$,
$F_\pi(\mu = 0)=86.4\,\mbox{MeV}$ estimated in the chiral 
limit~\cite{Gas:84,Gas:85a,Gas:85b}~\footnote{
 In Ref.~\cite{HYc}, it was assumed that 
 the scale dependent $F_\pi (\mu)$
 agrees with the scale-independent parameter $F_\pi$ 
 in the ChPT at $\mu = 0$ to obtain 
 $F_\pi(\mu=0)=86.4\pm9.7\,$MeV.
 Here, we simply
 use this value to get a rough estimate of the value of the
 parameter $a(0)$ as done in Ref.~\cite{HYc}.
} and
$F_\sigma^2(M_\rho)/F_\pi^2(0) = 2.03$ obtained
through the Wilsonian matching 
for $\Lambda_{\rm QCD}=400\,\mbox{MeV}$ and 
the matching scale $\Lambda=1.1\,\mbox{GeV}$
in Ref.~\cite{HYc},
we estimate the value of $a(0)$ at zero temperature as
\begin{equation}
 a(0) \simeq 2.31 \ .
\label{a0 val}
\end{equation}
This implies that the VD is well satisfied at $T=0$ even though
the value of the parameter $a$ at the scale $M_\rho$ is 
close to one~\cite{Harada:2001rf}.

Now, let us study
the direct $\gamma\pi\pi$ coupling in hot matter.
In general,
the electric mode and the magnetic mode of the photon
couple to the pions differently in hot matter,
so that there are two parameters as extensions of the parameter $a$.
Similarly to the one obtained at $T=0$ in Eq.~(\ref{gpp T0 0}),
in the low energy limit
the direct $\gamma\pi\pi$ interaction derived from
$\overline{\cal A}_\mu$-$\overline{\cal A}_\nu$ and
$\overline{\cal V}_\mu$-$\overline{\cal V}_\nu$ two-point
functions is expressed as
\begin{eqnarray}
&&
\Gamma_{\gamma\pi\pi}^{\mu}(p;q,k)
\nonumber\\
&& \quad
=
\frac{1}{\widetilde{F}(\bar{q};T)\widetilde{F}(\bar{k};T)}
\Biggl[
  q_\nu \, 
  \Pi_\perp^{\mu\nu}(q_0,\bar{q};T)
  -
  k_\nu \, 
  \Pi_\perp^{\mu\nu}(k_0,\bar{k};T)
\nonumber\\
&& \qquad\qquad\qquad\qquad\qquad
  {}- \frac{1}{2} (q-k)_\nu \,
    \Pi_\parallel^{\mu\nu}(p_0,\bar{p};T)
\Biggr]
\ ,
\label{direct g pi pi 0}
\end{eqnarray}
where $q$ and $k$ denote outgoing
momenta of the pions and $p = (q+k)$ is the photon momentum.
$\widetilde{F}$ is the wave function renormalization of 
the background 
$\bar{\pi}$ field given in Eq.~(\ref{pi wave renorm}).
Note that each pion is on its mass shell, so that
$q_0 = v_\pi(\bar{q}) \bar{q}$ and 
$k_0 = v_\pi(\bar{k}) \bar{k}$.
To define extensions of the parameter $a$, 
we consider the soft limit of the photon:
$p_0 \to 0$ and $\bar{p} \to 0$.~\footnote{%
  Note that we take the $p_0 \rightarrow 0$ limit first and then take
  the $\bar{p}\rightarrow0$ limit for definiteness.
  We think that this is natural since the form factor in the vacuum
  is defined for space-like momentum of the photon.
  }
Then the pion momenta become
\begin{equation}
 q_0 = - k_0 \ , \quad  \bar{q}=- \bar{k} \ .
\end{equation}
Note that while
only two components $\Pi_\perp^t$ and $\Pi_\perp^s$ appear
in $q_\nu \, \Pi_\perp^{\mu\nu}$
or $k_\nu \, \Pi_\perp^{\mu\nu}$,
$(q-k)_\nu \, \Pi_\parallel^{\mu\nu}$
includes all four components $\Pi_\parallel^t$,
$\Pi_\parallel^s$, $\Pi_\parallel^L$ and $\Pi_\parallel^T$.
Since the tree part of $\Pi_\parallel^L$ and $\Pi_\parallel^T$ is
$-2z_2\,p^2$ which vanishes at $p^2=0$, 
it is natural to use
only $\Pi_\parallel^t$ and $\Pi_\parallel^s$ to define the extensions
of the parameter $a$.
By including these two parts only, the temporal and
the spatial components of 
$\Gamma_{\gamma\pi\pi}^\mu$ are given by
\begin{eqnarray}
 \Gamma_{\gamma\pi\pi}^0(0;q,-q)
 &=& 
  \frac{2q_0}{\widetilde{F}^2(\bar{q};T)}
   \Bigl[ \Pi_\perp^t(q_0,\bar{q};T)-
                \frac{1}{2}\Pi_\parallel^t(0,0;T)
         \Bigr], \nonumber\\
 \Gamma_{\gamma\pi\pi}^i(0;q,-q)
 &=& 
  \frac{-2q_i}{\widetilde{F}^2(\bar{q};T)}
    \Bigl[ \Pi_\perp^s(q_0,\bar{q};T)-
                \frac{1}{2}\Pi_\parallel^s(0,0;T)
         \Bigr].
\end{eqnarray}
Thus we define $a^t(T)$ and $a^s(T)$ as
\begin{eqnarray}
 a^t(\bar{q};T)
 &=& \frac{\Pi_\parallel^t(0,0;T)}
         {\Pi_\perp^t(q_0,\bar{q};T)} \ , \nonumber\\
 a^s(\bar{q};T)
 &=& \frac{\Pi_\parallel^s(0,0;T)}
         {\Pi_\perp^s(q_0,\bar{q};T)} \ .
\end{eqnarray}
Here we should stress again that the pion momentum $q_\mu$ is
on mass-shell: $q_0 = v_\pi(\bar{q}) \bar{q}$.

In the HLS at one-loop level the above
$a^t(\bar{q};T)$ and $a^s(\bar{q};T)$
are expressed as
\begin{eqnarray}
 a^t (\bar{q};T)
 &=& a(0)\Biggl[ 1 + \frac{\overline{\Pi}_\parallel^t (0,0;T) -
     a(0)\overline{\Pi}_\perp^t (\bar{q},\bar{q};T)}
     {a(0) F_\pi^2(0;T)} \Biggr], \label{at expression}
\\
 a^s (\bar{q};T)
 &=& a(0)\Biggl[ 1 + \frac{\overline{\Pi}_\parallel^s (0,0;T) -
     a(0)\overline{\Pi}_\perp^s (\bar{q},\bar{q};T)}
     {a(0) F_\pi^2(0;T)} \Biggr], \label{as expression}
\end{eqnarray}
where $a(0)$ is defined in Eq.~(\ref{eq:Defa}).
{}From Eq.~(\ref{Pi bar V ts}) together with
Eq.~(\ref{B0B_static}) we obtain 
$\overline{\Pi }^{t}_\parallel$ and $\overline{\Pi }^{s}_\parallel$
in Eqs.~(\ref{at expression}) and (\ref{as expression}) as
\begin{eqnarray}
 \overline{\Pi }^{t}_\parallel(0,0;T)
 &=& \overline{\Pi }^{s}_\parallel(0,0;T) \nonumber\\
 &=& - \frac{N_f}{4}
   \Biggl[ 
     a^2 \tilde{I}_{2}(T) 
     - \tilde{J}_{1}^2(M_\rho;T) 
     + 2 \tilde{J}_{-1}^0(M_\rho;T) 
   \Biggr].
\label{PiVts zero limits}
\end{eqnarray}

Before going to the analysis near the critical temperature,
let us study the temperature dependence of the parameters
$a^t(\bar{q};T)$ and $a^s(\bar{q};T)$ in the low temperature
region.
At low temperature $T \ll M_\rho$
the functions $\tilde{J}_{1}^2(M_\rho;T)$ and
$\tilde{J}_{-1}^0(M_\rho;T)$ are suppressed by 
$e^{-M_\rho/T}$.
Then the dominant contribution is given by 
$\tilde{I}_{2}(T)=T^2/12$.
Combining this with Eq.~(\ref{t-low-T}) and (\ref{s-low-T}), we obtain
\begin{equation}
 a^t \simeq a^s \simeq
         a(0) \left[
         1 + \frac{N_f}{12} \left( 1 - \frac{a^2}{4 a(0)} \right)
         \frac{T^2}{F_\pi^2(0;T)} \right],
  \label{at as form low T}
 \end{equation}
where $a$ is the parameter renormalized at the scale $\mu=M_\rho$,
while $a(0)$ is defined in Eq.~(\ref{eq:Defa}).
We think that the intrinsic temperature dependences are small in the
low temperature region, so that we use the values of parameters at
$T=0$ to estimate the temperature dependent correction to the above
parameters.
By using $F_\pi(0)=86.4\,\mbox{MeV}$,
$a(0) \simeq 2.31$ given in Eq.~(\ref{a0 val}) and
$a(M_\rho) = 1.38$ 
obtained through the Wilsonian matching for
$(\Lambda_{\rm QCD}\,,\,\Lambda) = (0.4\,,\,1.1)\,\mbox{GeV}$
and $N_f = 3$~\cite{HYc},
$a^t$ and $a^s$ in Eq.~(\ref{at as form low T}) are evaluated as
\begin{equation}
a^t \simeq a^s \simeq
a(0) \left[
  1 + 0.066 \left( \frac{T}{50\,\mbox{MeV}} \right)^2 
\right]
\ .
\end{equation}
This implies that the parameters $a^t$ and $a^s$ increase with
temperature in the low temperature region.
However, since the correction is small, we conclude that the
vector dominance is well satisfied in the low temperature region.

At higher temperature
the intrinsic thermal effects become more important.
As we have shown in section~\ref{sec:ITE},
the parameters $(g,a)$ approach $(0,1)$
for $T \to T_c$ by the intrinsic temperature dependences,
and then the parametric vector meson mass $M_\rho$ vanishes.
As we have shown in section~\ref{sec:PDCPV},
near the critical temperature
$\Pi_\perp^t$ and $\Pi_\perp^s$
in Eqs.~(\ref{at expression}) and (\ref{as expression})
approach the following expressions:
\begin{eqnarray}
 \overline{\Pi}_\perp^t (\bar{q},\bar{q};T)
  \stackrel{T \to T_c}{\to} {}- \frac{N_f}{24}T^2 , \nonumber\\
 \overline{\Pi}_\perp^s (\bar{q},\bar{q};T)
  \stackrel{T \to T_c}{\to} {}- \frac{N_f}{24}T^2 .
\label{VM limits Pit Pis}
\end{eqnarray}
On the other hand, the functions
$\Pi_\parallel^{t}$ and
$\Pi_\parallel^{s}$ in Eq.~(\ref{PiVts zero limits})
at the limit of $M_\rho/T \rightarrow0$ and $a\rightarrow1$
become
\begin{equation}
 \bar{\Pi }^{t}_\parallel(0,0;T)
 = \bar{\Pi }^{s}_\parallel(0,0;T)
 \rightarrow - \frac{N_f}{2} \tilde{I}_{2}(T) 
 = - \frac{N_f}{24} T^2 \ .
\label{VM limit PiVts}
\end{equation}
Furthermore, from Eq.~(\ref{a0 expression}), the parameter
$a(0)$ approaches $1$ for $M_\rho\rightarrow0$ and
$F_\sigma^2(M_\rho)/F_\pi^2(0)\rightarrow1$:
\begin{equation}
a(0) \rightarrow 1 \ .
\label{VM limit a0}
\end{equation}
{}From the above limits in Eqs.~(\ref{VM limits Pit Pis}),
(\ref{VM limit PiVts})
and (\ref{VM limit a0}), the numerators of 
$a^t(\bar{q};T)$ and $a^s(\bar{q};T)$ 
in Eqs.~(\ref{at expression}) and (\ref{as expression})
behave as
\begin{eqnarray}
 \overline{\Pi}_\parallel^t (0,0;T) - a(0)\overline{\Pi}_\perp^t
 (\bar{q},\bar{q};T) \to 0, \nonumber\\
 \overline{\Pi}_\parallel^s (0,0;T) - a(0)\overline{\Pi}_\perp^s
 (\bar{q},\bar{q};T) \to 0.
\end{eqnarray}
Thus 
we obtain
 \begin{equation}
  a^t(\bar{q};T), a^s(\bar{q};T) 
   \stackrel{T \to T_c}{\to} 1 \ .
 \end{equation}
This implies that the vector dominance is largely
violated near the critical temperature.


\section{Summary and Discussions}
\label{sec:SD}

In this paper
we first
showed the detailed calculations of the two-point functions
in the background field gauge.
Then, we showed how to extend 
the Wilsonian matching to the version at non-zero temperature.
Throughout this paper, we assume that the chiral phase transition
is of second or weakly first order
so that the current correlators agree with each other 
at the critical temperature.
Thus the Wilsonian matching leads to the following conditions:
\begin{eqnarray}
 g(\Lambda;T) \stackrel{T \to T_c}{\to} 0, \qquad
 a(\Lambda;T) \stackrel{T \to T_c}{\to} 1. \label{VMCs}
\end{eqnarray}
It is important that these conditions for the bare parameters are
realized by the intrinsic thermal effect only,
and further $(g,a) = (0,1)$ is a fixed point of RGEs.
In order to be consistent with the chiral symmetry restoration in QCD,
the axial vector current correlator must agree with the vector current
correlator in the limit $T \to T_c$.
This agreement is satisfied when we
incorporate not only the hadronic corrections 
but also intrinsic effects into the physical quantities.

In section~\ref{sec:VMM}, 
we studied the vector meson pole mass including the hadronic
thermal corrections calculated in the background field gauge.
In the low temperature region $T \ll M_\rho$,
the vector meson mass increases as $T^4$
dominated by the pion loop.
When we take the limit $T \to T_c$,
it is crucial that the parameter $M_\rho$ goes to zero
by the intrinsic thermal effect.
Then we showed that the pole mass of the vector meson vanishes:
 \begin{eqnarray}
  m_\rho (T) \stackrel{T \to T_c}{\to} 0. \nonumber
 \end{eqnarray}
This implies that
the vector manifestation is realized in hot matter
as was first shown in Ref.~\cite{HSasaki} where we used the hadronic
thermal corrections calculated in the Landau gauge~\cite{HS}.
We should stress that the conditions in Eq.~(\ref{VMCs})
realized by the {\it intrinsic thermal effects},
which we call ``the VM conditions in hot matter'',
are essential for the VM to take place in hot matter.
One might think that the VM is same as 
Georgi's vector realization~\cite{Georgi},
in which the order parameter $f_\pi$ is non-zero 
although the chiral symmetry is unbroken.
However in the VM the order parameter certainly becomes zero
by the quadratic divergence
and chiral symmetry is restored by massless vector meson.
Therefore the VM is consistent with the Ward-Takahashi identity
since it is the Wigner realization~\cite{Yamawaki}.

In section~\ref{sec:PDCPV}, 
we studied the temperature dependence of the
pion velocity together with those of 
the temporal and spatial pion decay constants
in the low temperature region.
In hot matter, the pion velocity $v_\pi$ is not the speed of light
because of the lack of Lorentz invariance.
Actually we confirmed that
\begin{eqnarray}
 v_\pi^2 (\bar{p};T) \simeq 1 - \frac{N_f}{15}
   \frac{a \pi^2}{F_\pi^2 M_\rho^2} T^4\, < 1
 \qquad \mbox{for} \quad T \ll M_\rho. \nonumber
\end{eqnarray}
We briefly reviewed how the 
temporal and spatial pion decay constants
vanish at the critical point simultaneously
following Ref.~\cite{HKRS}.
Then we estimated the value of the critical temperature.
{}From the order parameter $f_\pi$ including the intrinsic thermal effect
as well as the hadronic correction,
we obtained the following values for several choices of
the matching scale ($\Lambda = 0.8\, \mbox{-}\, 1.1$\,GeV)
and the scale of QCD 
($\Lambda_{\rm QCD}= 0.30 \, \mbox{-} 0.45$\,GeV):
\begin{eqnarray*}
 T_c = 200\, \mbox{-}\, 250 \,\, \mbox{MeV}.
\end{eqnarray*}
We should note that the above values may be changed when we adopt
a different way to estimate the matrix elements of operators 
in the OPE side,
e.g., an estimation with the dilute pion gas approximation~\cite{HKL}
or that by the lattice QCD calculation~\cite{lattice}. 
Even when we choose one way to estimate the matrix elements in the OPE, 
some temperature effects are supposed to be left out 
due to the truncation of the OPE to neglect higher order operators,
inclusion of which will cause a small change of the above values
of the critical temperature.
The important point is that as a result of the Wilsonian matching,
$T_c$ is obtained as in Eq.~(\ref{eq:T_c}) 
in terms of the quark and gluonic condensates, not hadronic degrees of
freedom.   It is expected that the value of $T_c$ may become smaller
than that obtained in this paper by including the higher order
corrections.

Finally in section~\ref{sec:PaVVD}, we presented a new prediction 
associated with the validity of vector dominance (VD) in hot matter.
In the HLS model at zero temperature, 
the Wilsonian matching predicts $a \simeq2$~\cite{HYa,HYc}
which guarantees the VD of the
electromagnetic 
form factor of the pion.
Even at non-zero temperature,
this is valid as long as we consider the thermal effects
in the low temperature region,
where the intrinsic temperature dependences are negligible.
We showed that, as a consequence of including the intrinsic effect,
the VD is largely violated at the critical temperature:
\begin{eqnarray*}
 a^t(\bar{p};T) \stackrel{T \to T_c}{\to} 1, \qquad
 a^s(\bar{p};T) \stackrel{T \to T_c}{\to} 1.
\end{eqnarray*}
In general, full temperature dependences include both hadronic and
intrinsic thermal effects.
Then there exist the violations of VD and universality of
the $\rho$-coupling at generic temperature,
although at low temperature the VD and universality are 
approximately satisfied.

In several analyses such as the one on the dilepton spectra 
in hot matter
done
in Ref~\cite{Rapp-Wambach:00}, the VD is assumed to be held even in
the high temperature region.
Our result indicates that the assumption of 
the VD may need to be weakened, at least by some amount,
for consistently including the
effect of the dropping mass of the vector meson into the analysis.

Several comments are in order:

The unitarity of the scalar channel
of $\pi$-$\pi$ scattering will be broken around 400-500 MeV in the
present model although the existence of the $\rho$ meson helps
a little, as was studied in Ref.~\cite{Harada:1995dc}.
The unitarity of the vector sector, on the other hand, is not
violated due to the existence of the vector mesons.
Then, we do not consider the scalar sector in the present 
analysis,
and
we restrict ourselves to the quantities related to the vector and
axial vector sector.
In Ref.~\cite{HYc}, it was discussed 
that in the limit approaching the critical
point, 
the scalar meson may decouple from the theory since the scalar meson 
belongs to the different chiral representation from both the pion and 
the longitudinal vector meson.

In section~\ref{sec:VMM},
we studied the pole mass of the vector meson at the critical temperature.
It is also important to study the screening mass of the vector meson
which is determined from the vector current correlator
as done in the lattice calculation.
It is not necessary that the correlation function of vector current be
divergent at the critical temperature since 
$\langle \bar{q}\gamma^\mu q \rangle$ is not the order parameter
of the spontaneous chiral symmetry breaking.
The vector susceptibility $\chi_V$ is related to 
the correlation length of the vector current in the static limit
and is in fact finite at the critical temperature
as studied in Ref.~\cite{HKRS}, which is consistent with the result
by the lattice QCD calculation~\cite{Gottlieb:ac,Allton:2002zi}.
This may imply that the screening mass of the vector meson is finite
at the critical temperature.

In the present analysis we neglected the Lorentz symmetry violating
effects at bare level since they are small as was shown 
in Ref.~\cite{HKRS} (see also section~\ref{sec:ITE}).
It is interesting to include such small corrections at bare level 
and study their effects to the physical quantities 
(see e.g., Refs.~\cite{Sasaki:2003qj,Harada:2003at}).

It is important to study which universality class the VM belongs to. 
The analysis of critical exponents gives the answer of this problem.
We will study this in a future publication. 

In this paper, we studied
the physical quantities including both hadronic and intrinsic 
thermal effects at the limit $T \to T_c$ as well as 
in the low temperature region,
$T \ll M_\rho$.
It is important to study the behavior away from $T_c$
since the properties of vector mesons in the temperature region
from 100 MeV to 200 MeV are needed for the analysis of RHIC data.
This will be done in future works.


\section*{Acknowledgment}

We would like to thank Doctor Youngman Kim, Professor Mannque Rho 
and Professor Koichi Yamawaki for useful discussions and comments.
We are grateful to Professor Gerry Brown for critical reading of
the manuscript.
MH would like to thank Professor Dong-Pil Min for his hospitality
during the stay in Seoul National University where part of this work
was done.
The work of MH was supported in part by the
Brain Pool program (\#012-1-44) provided by the Korean Federation
of Science and Technology Societies.
This work is supported in part by the 21st Century COE
Program of Nagoya University provided by Japan Society for the
Promotion of Science (15COEG01).

\newpage

\appendix

\begin{flushleft}
\bf\LARGE
Appendices
\end{flushleft}

\setcounter{section}{0}
\renewcommand{\thesection}{\Alph{section}}
\setcounter{equation}{0}
\renewcommand{\theequation}{\Alph{section}.\arabic{equation}}

\section{Polarization Tensors at Finite Temperature}
\label{app:PTFT}

At non-zero temperature
there exist four independent polarization tensors,
$u^\mu u^\nu, (g^{\mu\nu}-u^\mu u^\nu),\, P_L^{\mu\nu}$
and $P_T^{\mu\nu}$.
The rest frame of medium is shown by $u^\mu=(1,\vec{0})$.
$P_L^{\mu\nu}$ and $P_T^{\mu\nu}$ are given by~\cite{Kap}
  \begin{eqnarray}
   P_{T \mu \nu} &=& g_{\mu i}\Bigl( \delta _{ij} -
                     \frac{\vec{p_i}\vec{p_j}}{\bar{p}^2} \Bigr)
                     g_{j \nu}, \nonumber \\
   P_{L \mu \nu} &=& -\Bigl( g_{\mu \nu} - \frac{p_\mu p_\nu}{p^2} \Bigr)-
                     P_{T \mu \nu}, 
  \label{A.1}
  \end{eqnarray}
where we 
define $p^\mu = (p_0\,,\,\vec{p})$ and 
$\bar{p}=|\vec{p}|$.
They have the following properties:
\begin{eqnarray}
 &P_{L\mu}^\mu = -1,\, P_{T\mu}^\mu = -2,& \nonumber\\
 &P_{L\mu\alpha}P_L^{\alpha\nu} = -{P_{L\mu}}^\nu,\,
 P_{T\mu\alpha}P_T^{\alpha\nu} = -{P_{T\mu}}^\nu,& \nonumber\\
 &P_{L\mu\alpha}P_T^{\alpha\nu}  = 0.& 
\end{eqnarray}

Let us decompose a tensor $\Pi^{\mu\nu}(p_0,\bar{p})$ into
\begin{equation}
 \Pi^{\mu\nu}=u^\mu u^\nu \Pi^t +
              (g^{\mu\nu}-u^\mu u^\nu)\Pi^s +
              P_L^{\mu\nu}\Pi^L + P_T^{\mu\nu}\Pi^T. 
\end{equation}
Each component is obtained as
\begin{eqnarray}
 \Pi^t &=& \Pi^{00} + \frac{\vec{p}_i}{p_0}\Pi^{i0}, \nonumber\\
 \Pi^s &=& -\frac{\vec{p}_i}{\bar{p}}\Pi^{ij}\frac{\vec{p}_j}{\bar{p}} - 
          \frac{p_0 \vec{p}_i}{\bar{p}^2}\Pi^{i0}, \nonumber\\
 \Pi^L-\Pi^s &=& \frac{\vec{p}_i}{\bar{p}}\Pi^{ij}\frac{\vec{p}_j}{\bar{p}}+ 
                \frac{\vec{p}_i}{p_0}\Pi^{i0}, \nonumber\\
 \Pi^T-\Pi^s &=& \frac{1}{2}P_{T\mu\nu}\Pi^{\mu\nu}. 
\end{eqnarray}


\section{Loop Integrals at Finite Temperature}
\label{app:LIFT}

In this appendix we list the explicit forms of the
functions appearing in the hadronic thermal corrections,
$\overline{A}_{0}$, $\overline{B}_{0}$ and
$\overline{B}^{\mu\nu}$ 
[see Eqs.~(\ref{def:A0 2})--(\ref{def:Bmunu 11}) for definitions]
in various limits relevant to the
present analysis.

The functions $\overline{A}_{0}$,
which are independent of
external momentum $p_\mu$,
is given by
 \begin{eqnarray}
  \overline{A}_{0}(M_\rho ;T)
     &=& \tilde{J}^2_{1}(M_\rho ;T)\ , 
  \label{A0BM} \\
  \overline{A}_{0}(0;T)
     &=& \tilde{I}_{2}(T)\ ,
  \label{A0B0}
 \end{eqnarray}
where $\tilde{J}^2_1$ and $\tilde{I}_2$ are 
defined 
in Eqs.~(\ref{I fun}) and (\ref{J fun}).

We list the relevant limits of the functions 
$\overline{B}_{0}(p_0,\bar{p};M_\rho,0;T)$ 
and $\overline{B}^{\mu\nu}(p_0,\bar{p};M_\rho,0;T)$ 
which appear in the two-point function
of $\overline{\cal A}_\mu$.
When 
the pion momentum is taken as its on-shell, 
the function $\overline{B}_{0}$ becomes
\begin{eqnarray}
 &&\overline{B}_{0}
 (p_0 = \bar{p}+i\epsilon, \bar{p};M_\rho,0;T) \nonumber\\
 &=& \int\,\frac{d^3 k}{(2\pi)^3}
   \Biggl[ \frac{-1}{2\omega_\rho}\frac{1}{e^{\omega_\rho/T}-1}
    \Biggl\{ \frac{1}{(\omega_\rho - \bar{p})^2 - (\omega_\pi^p)^2 -
                       2i\epsilon (\omega_\rho - \bar{p})} +
             \frac{1}{(\omega_\rho + \bar{p})^2 - (\omega_\pi^p)^2 +
                       2i\epsilon (\omega_\rho + \bar{p})}
    \Biggr\}\nonumber\\ 
  &&{}+
           \frac{-1}{2\omega_\pi^p}\frac{1}{e^{\omega_\pi^p/T}-1}
    \Biggl\{ \frac{1}{(\omega_\pi^p - \bar{p})^2 - \omega_\rho^2 -
                       2i\epsilon (\omega_\pi^p - \bar{p})} +
             \frac{1}{(\omega_\pi^p + \bar{p})^2 - \omega_\rho^2 +
                       2i\epsilon (\omega_\pi^p + \bar{p})}
    \Biggr\}
   \Biggr]\ , \nonumber\\
\label{B0 A on-shell}
\end{eqnarray}
where we put $\epsilon \rightarrow +0$ to make the analytic
continuation of the frequency $p_0=i2\pi n T$ to the Minkowski
variable, and 
we defined 
\begin{equation}
\omega_\rho = \sqrt{|\vec{k}|^2 + M_\rho^2} \ ,
\quad
\omega_\pi^p = |\vec{k}-\vec{p}| \ .
\label{def: omega-rho omega-pi}
\end{equation}
Two components $\overline{B}^t$ and $\overline{B}^s$ 
of $\overline{B}^{\mu\nu}$ 
in the same limit are given by
\begin{eqnarray}
 &&\overline{B}^t
  (p_0 = \bar{p}+i\epsilon,\bar{p};M_\rho,0;T) \nonumber\\
 &=& \int\,\frac{d^3 k}{(2\pi)^3}
   \Biggl[ \frac{-1}{2\omega_\rho}\frac{1}{e^{\omega_\rho/T}-1}
    \Biggl\{ \frac{(2\omega_\rho - \bar{p})^2}
                   {(\omega_\rho - \bar{p})^2 - (\omega_\pi^p)^2 -
                     2i\epsilon (\omega_\rho - \bar{p})} +
              \frac{(2\omega_\rho + \bar{p})^2}
                   {(\omega_\rho + \bar{p})^2 - (\omega_\pi^p)^2 +
                     2i\epsilon (\omega_\rho + \bar{p})} \nonumber\\
  &&{}-\frac{\vec{p}\cdot(2\vec{k}-\vec{p})}{\bar{p}}
             \Biggl( \frac{2\omega_\rho - \bar{p}}
                           {(\omega_\rho - \bar{p})^2 - (\omega_\pi^p)^2 -
                             2i\epsilon (\omega_\rho - \bar{p})} -
                      \frac{2\omega_\rho + \bar{p}}
                           {(\omega_\rho + \bar{p})^2 - (\omega_\pi^p)^2 +
                             2i\epsilon (\omega_\rho + \bar{p})}
              \Biggr)
    \Biggr\} \nonumber\\
  &&{}+\frac{-1}{2\omega_\pi^p}\frac{1}{e^{\omega_\pi^p/T}-1}
    \Biggl\{ \frac{(2\omega_\pi^p - \bar{p})^2}
                   {(\omega_\pi^p - \bar{p})^2 - \omega_\rho^2 -
                     2i\epsilon (\omega_\pi^p - \bar{p})} +
              \frac{(2\omega_\pi^p + \bar{p})^2}
                   {(\omega_\pi^p + \bar{p})^2 - \omega_\rho^2 +
                     2i\epsilon (\omega_\pi^p + \bar{p})} \nonumber\\
   &&{}+\frac{\vec{p}\cdot(2\vec{k}-\vec{p})}{\bar{p}}
             \Biggl( \frac{2\omega_\pi^p - \bar{p}}
                           {(\omega_\pi^p - \bar{p})^2 - \omega_\rho^2 -
                             2i\epsilon (\omega_\pi^p - \bar{p})} -
                      \frac{2\omega_\pi^p + \bar{p}}
                           {(\omega_\pi^p + \bar{p})^2 - \omega_\rho^2 +
                             2i\epsilon (\omega_\pi^p + \bar{p})}
              \Biggr)
    \Biggr\}
   \Biggr]\ ,
\nonumber\\
\label{Bt A on-shell}
\\
 &&\overline{B}^s
  (p_0 = \bar{p}+i\epsilon,\bar{p};M_\rho,0;T) \nonumber\\
 &=& \int\,\frac{d^3 k}{(2\pi)^3}
   \Biggl[ \frac{1}{2\omega_\rho}
           \frac{1}{e^{\omega_\rho/T}-1} \nonumber\\ 
 &&\times
    \Biggl\{ \frac{(2\vec{k}\cdot\vec{p}-\bar{p}^2)^2}{\bar{p}^2}
     \Biggl( \frac{1}{(\omega_\rho - \bar{p})^2 - (\omega_\pi^p)^2 -
                       2i\epsilon (\omega_\rho - \bar{p})} +
              \frac{1}{(\omega_\rho + \bar{p})^2 - (\omega_\pi^p)^2 +
                       2i\epsilon (\omega_\rho + \bar{p})}
     \Biggr) \nonumber\\ 
  &&{}-\frac{\vec{p}\cdot (2\vec{k}-\vec{p})}{\bar{p}}
     \Biggl( \frac{2\omega_\rho - \bar{p}}
                   {(\omega_\rho - \bar{p})^2 - (\omega_\pi^p)^2 -
                      2i\epsilon (\omega_\rho - \bar{p})}-
              \frac{2\omega_\rho + \bar{p}}
                   {(\omega_\rho + \bar{p})^2 - (\omega_\pi^p)^2 +
                     2i\epsilon (\omega_\rho + \bar{p})}
     \Biggr)
    \Biggr\} \nonumber\\
  &&\qquad\quad{}+\frac{1}{2\omega_\pi^p}
           \frac{1}{e^{\omega_\pi^p/T}-1} \nonumber\\ 
  &&\times
    \Biggl\{ \frac{(2\vec{k}\cdot\vec{p}-\bar{p}^2)^2}{\bar{p}^2}
     \Biggl( \frac{1}{(\omega_\pi^p - \bar{p})^2 - \omega_\rho^2 -
                        2i\epsilon (\omega_\pi^p - \bar{p})} +
              \frac{1}{(\omega_\pi^p + \bar{p})^2 - \omega_\rho^2 +
                        2i\epsilon (\omega_\pi^p + \bar{p})}
     \Biggr) \nonumber\\ 
  &&{}+\frac{\vec{p}\cdot (2\vec{k}-\vec{p})}{\bar{p}}
     \Biggl( \frac{2\omega_\pi^p - \bar{p}}
                   {(\omega_\pi^p - \bar{p})^2 - \omega_\rho^2 -
                     2i\epsilon (\omega_\pi^p - \bar{p})}-
              \frac{2\omega_\pi^p + \bar{p}}
                   {(\omega_\pi^p + \bar{p})^2 - \omega_\rho^2 +
                     2i\epsilon (\omega_\pi^p + \bar{p})}
     \Biggr)
    \Biggr\}
   \Biggr]\ . 
\nonumber\\
\label{Bs A on-shell}
\end{eqnarray}
We further take the $M_\rho\rightarrow0$ limits of the above
expressions.
The limit of 
$\overline{B}_{0}(p_0 = \bar{p}+i\epsilon, \bar{p};M_\rho,0;T)$
includes the infrared logarithmic divergence $\ln M_\rho^2$.
However, it appears multiplied by $M_\rho^2$ in 
$\overline{\Pi}_\perp^t$ and $\overline{\Pi}_\perp^s$, and
the product
$M_\rho^2
\overline{B}_{0}(p_0 = \bar{p}+i\epsilon, \bar{p};M_\rho,0;T)$
vanishes at $M_\rho\rightarrow0$ limit:
\begin{equation}
\lim_{M_\rho \rightarrow 0}
M_\rho^2
\overline{B}_{0}(p_0 = \bar{p}+i\epsilon, \bar{p};M_\rho,0;T)
= 0
\ .
\label{B0 A on-shell VM}
\end{equation}
As for
$\overline{B}^{t,s}(p_0=\bar{p}+i\epsilon,\bar{p};M_\rho,0;T)$,
we obtain
\begin{eqnarray}
 \lim_{M_\rho \to 0}
 \overline{B}^t(p_0=\bar{p}+i\epsilon,\bar{p};M_\rho,0;T)
 &=& -2 \tilde{I}_{2}(T)\ , \nonumber\\
 \lim_{M_\rho \to 0}
 \overline{B}^s(p_0=\bar{p}+i\epsilon,\bar{p};M_\rho,0;T)
 &=& -2 \tilde{I}_{2}(T)\ .
\label{Bts A on-shell VM}
\end{eqnarray}

In the static limit ($p_0 \to 0$), the functions
$M_\rho^2 \overline{B}_{0}(p_0,\bar{p};M_\rho,0;T)$ and
$\overline{B}^t(p_0,\bar{p};M_\rho,0;T)-
\overline{B}^L(p_0,\bar{p};M_\rho,0;T)$ become
\begin{eqnarray}
 &&\lim_{p_0 \to 0}
 M_\rho^2 \overline{B}_{0}(p_0,\bar{p};M_\rho,0;T) \nonumber\\
 &&\qquad\qquad = M_\rho^2 \int\frac{d^3 k}{(2\pi)^3}
   \frac{-1}{\omega_\rho^2 - (\omega_\pi^p)^2}
   \Biggl[ \frac{1}{\omega_\rho}\frac{1}{e^{\omega_\rho/T}-1} -
           \frac{1}{\omega_\pi^p}\frac{1}{e^{\omega_\pi^p/T}-1}
   \Biggr],
\label{B0B_static} \\
 &&\lim_{p_0 \to 0}
 \Bigl[
  \overline{B}^t(p_0,\bar{p};M_\rho,0;T) -
  \overline{B}^L(p_0,\bar{p};M_\rho,0;T)
 \Bigr] \nonumber\\
 &&\qquad\qquad = \int\frac{d^3 k}{(2\pi)^3}
   \frac{-4}{\omega_\rho^2 - (\omega_\pi^p)^2}
   \Biggl[ \frac{\omega_\rho}{e^{\omega_\rho/T}-1} -
           \frac{\omega_\pi^p}{e^{\omega_\pi^p/T}-1}
   \Biggr].
\label{Bt-BL static}
\end{eqnarray}
Taking the low momentum limits of these expressions, we obtain
\begin{eqnarray}
 &&\lim_{\bar{p} \to 0}
 M_\rho^2 \overline{B}_{0}(p_0=0,\bar{p};M_\rho,0;T)
  = \tilde{I}_{2}(T) - \tilde{J}_{1}^2(M_\rho;T), 
\label{B0B_staticL} \\
 &&\lim_{\bar{p} \to 0}
 \Bigl[
  \overline{B}^t(p_0=0,\bar{p};M_\rho,0;T) -
  \overline{B}^L(p_0=0,\bar{p};M_\rho,0;T)
 \Bigr] \nonumber\\
 &&\qquad\qquad = -\frac{4}{M_\rho^2}\Bigl[
     \tilde{J}_{-1}^2(M_\rho;T) - \tilde{I}_{4}(T) \Bigr].
\label{Bt-BL staticL}
\end{eqnarray}
Moreover, in the VM limit ($M_\rho \to 0$), these are reduced to
\begin{eqnarray}
 &&\lim_{\bar{p} \to 0}
 M_\rho^2 \overline{B}_{0}(p_0=0,\bar{p};M_\rho,0;T)
 \stackrel{M_\rho \to 0}{\to} 0, \nonumber\\
 &&\lim_{\bar{p} \to 0}
 \Bigl[
  \overline{B}^t(p_0=0,\bar{p};M_\rho,0;T) -
  \overline{B}^L(p_0=0,\bar{p};M_\rho,0;T)
 \Bigr]
 \stackrel{M_\rho \to 0}{\to} 2\tilde{I}_{2}(T)\ .
\label{B0B Bt-BL staticLM}
\end{eqnarray}

We consider the functions $\overline{B}_0$ and 
$\overline{B}^{\mu\nu}$ appearing in the two-point functions of
$\overline{\cal V}_\mu$ and $\overline{V}_\mu$.
$\overline{B}^t$ and $\overline{B}^s$ are
calculated as
\begin{eqnarray}
 &\overline{B}^{t}(p_0,\bar{p};0,0;T)
 =\overline{B}^{s}(p_0,\bar{p};0,0;T)
 = - 2 \overline{A}_{0}(0;T)
 = -2 \tilde{I}_{2}(T),& \nonumber\\
 &\overline{B}^{t}(p_0,\bar{p};M_\rho ,M_\rho ;T)
 = \overline{B}^{s}(p_0,\bar{p};M_\rho ,M_\rho ;T)
 = - 2 \overline{A}_{0}(M_\rho;T)
 = -2 \tilde{J}_{1}^2(M_\rho;T).&
\nonumber\\ 
\label{B.8}
\end{eqnarray}

The function $\overline{B}_{0}$ is expressed as
\begin{eqnarray}
 &&\overline{B}_{0}(p_0,\bar{p};M_\rho,M_\rho;T) \nonumber\\
 &=& \int\,\frac{d^3 k}{(2\pi)^3}
   \Biggl[ \frac{-1}{2\omega_\rho}\frac{1}{e^{\omega_\rho/T}-1}
    \Biggl\{ \frac{1}{(\omega_\rho - p_0)^2 - (\omega_\rho^p)^2} +
             \frac{1}{(\omega_\rho + p_0)^2 - (\omega_\rho^p)^2}
    \Biggr\}\nonumber\\ 
 &&\qquad\quad {}+
           \frac{-1}{2\omega_\rho^p}\frac{1}{e^{\omega_\rho^p/T}-1}
    \Biggl\{ \frac{1}{(\omega_\rho^p - p_0)^2 - \omega_\rho^2} +
             \frac{1}{(\omega_\rho^p + p_0)^2 - \omega_\rho^2}
    \Biggr\}
   \Biggr]\ , 
\end{eqnarray}
where we define 
\begin{equation}
\omega_\rho^p = \sqrt{ |\vec{k}-\vec{p}|^2 + M_\rho^2 }
\ .
\end{equation}
In the static limit $(p_0 \to 0)$, the functions $\overline{B}_0$ and
$\overline{B}^L$ are expressed as
\begin{eqnarray}
 &&\lim_{p_0 \to 0}
 M_\rho^2 \overline{B}_{0}(p_0,\bar{p};M_\rho ,M_\rho ;T)
\nonumber\\
 &&\qquad\qquad = M_\rho^2 \int \frac{d^3 k}{(2\pi)^3}
   \frac{-1}{\omega_\rho^2 - (\omega_\rho^p)^2}
   \Biggl[ \frac{1}{\omega_\rho}\frac{1}{e^{\omega_\rho/T}-1} -
           \frac{1}{\omega_\rho^p}\frac{1}{e^{\omega_\rho^p/T}-1}
   \Biggr], \\
 &&\lim_{p_0 \to 0}
 \overline{B}^L(p_0,\bar{p};M_\rho,M_\rho;T) \nonumber\\
 &&\qquad\qquad= \int \frac{d^3 k}{(2\pi)^3}
   \frac{1}{\vec{p}\cdot(2\vec{k}-\vec{p})}
   \Biggl[ \frac{1}{\omega_\rho}\frac{1}{e^{\omega_\rho/T}-1}
    \Bigl\{ 4\omega_\rho^2 - \vec{p}\cdot(2\vec{k}-\vec{p})
    \Bigr\} \nonumber\\
 &&\qquad\qquad\qquad\qquad\qquad\qquad\quad
      {}-  \frac{1}{\omega_\rho^p}\frac{1}{e^{\omega_\rho^p/T}-1}
    \Bigl\{ 4(\omega_\rho^p)^2 + \vec{p}\cdot(2\vec{k}-\vec{p})
    \Bigr\}
   \Biggr].
\end{eqnarray}
Taking the low momentum limit $(\bar{p} \to 0)$, 
these expressions become
\begin{eqnarray}
 &&\lim_{\bar{p} \to 0}
  M_\rho^2 \overline{B}_{0}(p_0=0,\bar{p};M_\rho ,M_\rho ;T)
  = \frac{1}{2}\Bigl[ \tilde{J}_{-1}^0(M_\rho;T) -
    \tilde{J}_{1}^2(M_\rho;T) \Bigr], \\
 &&\lim_{\bar{p} \to 0}
  \overline{B}^L(p_0=0,\bar{p};M_\rho,M_\rho;T)
  = {}-2\Bigl[ M_\rho^2 \tilde{J}_{1}^0(M_\rho;T) +
     2\tilde{J}_{1}^2(M_\rho;T) \Bigr].
\end{eqnarray}
In the VM limit $(M_\rho \to 0)$, these are reduced to
\begin{eqnarray}
 &&\lim_{\bar{p} \to 0}
  M_\rho^2 \overline{B}_{0}(p_0=0,\bar{p};M_\rho ,M_\rho ;T)
  \stackrel{M_\rho \to 0}{\to} 0, \\
 &&\lim_{\bar{p} \to 0}
  \overline{B}^L(p_0=0,\bar{p};M_\rho,M_\rho;T)
  \stackrel{M_\rho \to 0}{\to} {}-4\tilde{I}_2(T).
\end{eqnarray}

On the other hand, the functions 
$\overline{B}_0, \overline{B}^L-\overline{B}^s$
and $\overline{B}^T-\overline{B}^s$ in the low momentum limit $(\bar{p} \to 0)$
are given by
\begin{eqnarray}
 &&\lim_{\bar{p} \to 0}
 \overline{B}_{0}(p_0,\bar{p};M_\rho ,M_\rho ;T) 
 = \frac{1}{2}\tilde{F}^2_{3}(p_0;M_\rho ;T) \ ,
\label{B0B rest} \\
 &&\lim_{\bar{p} \to 0}\Bigl[
 \overline{B}^L(p_0,\bar{p};M_\rho,M_\rho;T)-
 \overline{B}^s(p_0,\bar{p};M_\rho,M_\rho;T)
 \Bigr]
  = \frac{2}{3}\tilde{F}_{3}^4(p_0;M_\rho;T), 
\label{BL rest}\\
 &&\lim_{\bar{p} \to 0}\Bigl[
 \overline{B}^T(p_0,\bar{p};M_\rho,M_\rho;T)-
 \overline{B}^s(p_0,\bar{p};M_\rho,M_\rho;T)
 \Bigr]
 = \frac{2}{3}\tilde{F}_{3}^4(p_0;M_\rho;T), 
  \label{BT rest}
\end{eqnarray}
where the function $\tilde{F}_{3}^n$ is
defined in Appendix D.


\section{\label{app:C}
  Quantum Corrections}

In this appendix we briefly summarize the quantum corrections to 
the two-point functions.

Let us define
the functions $B_0^{\rm(vac)}$, $B^{{\rm(vac)}\mu\nu}$ and
$A_0^{\rm(vac)}$
by the following integrals~\cite{HYc}:
\begin{eqnarray}
  A_0^{\rm(vac)}(M)
   &=& \int \frac{d^n k}{i(2\pi)^4}
      \frac{1}{M^2 - k^2}\ , 
\label{A0vac def}
\\
  B_0^{\rm(vac)}(p;M_1,M_2)
   &=& \int \frac{d^n k}{i(2\pi)^4}
      \frac{1}{[M_1^2-k^2][M_2^2-(k-p)^2]}\ , 
\label{B0vac def}
\\
  B^{{\rm(vac)}\mu\nu}(p;M_1,M_2)
   &=& \int \frac{d^n k}{i(2\pi)^4}
      \frac{(2k-p)^\mu (2k-p)^\nu }{[M_1^2-k^2][M_2^2-(k-p)^2]}\ .
\label{Bmnvac def}
\end{eqnarray}
In the present analysis it is important to include the quadratic
divergences to obtain the RGEs in the Wilsonian sense.
Here, following 
Refs.~\cite{HY:conformal,HYa,HYc}, we adopt the
dimensional regularization and identify the quadratic divergences with
the presence of poles of ultraviolet origin at $n=2$~\cite{Veltman}.
This can be done by the following replacement in the Feynman
integrals:
\begin{equation}
\int \frac{d^n k}{i (2\pi)^n} \frac{1}{-k^2} \rightarrow 
\frac{\Lambda^2} {(4\pi)^2} \ ,
\qquad
\int \frac{d^n k}{i (2\pi)^n} 
\frac{k_\mu k_\nu}{\left[-k^2\right]^2} \rightarrow 
- \frac{\Lambda^2} {2(4\pi)^2} g_{\mu\nu} \ .
\label{quad repl}
\end{equation}
On the other hand, 
the logarithmic divergence is identified with the pole at 
$n=4$:
\begin{equation}
\frac{1}{\bar{\epsilon}} + 1 \rightarrow
\ln \Lambda^2
\ ,
\label{logrepl:2}
\end{equation}
where
\begin{equation}
\frac{1}{\bar{\epsilon}} \equiv
\frac{2}{4 - n } - \gamma_E + \ln (4\pi)
\ ,
\end{equation}
with $\gamma_E$ being the Euler constant.

By using the replacements in Eqs.~(\ref{quad repl}) and
(\ref{logrepl:2}), the integrals in
Eqs.~(\ref{A0vac def}), (\ref{B0vac def}) and (\ref{Bmnvac def})
are evaluated as
\begin{eqnarray}
&&
A_0^{\rm(vac)}(M) 
=
\frac{\Lambda^2} {(4\pi)^2}
- \frac{M^2}{(4\pi)^2} \ln \frac{\Lambda^2}{M^2}
\ ,
\label{A0vac}
\\
&&
  B_0^{\rm(vac)}(p^2;M_1,M_2) 
=
\frac{1}{(4\pi)^2}
\left[
  \ln \Lambda^2 - 1 - F_0(p^2;M_1,M_2)
\right]
\ ,
\label{B0vac}
\\
&&
  B^{{\rm(vac)}\mu\nu}(p;M_1,M_2)
\nonumber\\
&& \quad
=
  - g^{\mu\nu} \frac{1}{(4\pi)^2}
  \left[ 
    2 \Lambda^2 - M_1^2 \ln \frac{\Lambda^2}{M_1^2}
   - M_2^2 \ln \frac{\Lambda^2}{M_2^2}
   - (M_1^2-M_2^2) F_A(p^2;M_1,M_2)
  \right]
\nonumber\\
&& \qquad
  {}- \left( g^{\mu\nu}p^2 - p^\mu p^\nu \right) 
    \frac{1}{(4\pi)^2} 
  \left[ 
    \frac{1}{3} \ln \Lambda^2 - F_0(p^2;M_1,M_2)
    + 4 F_3(p^2;M_1,M_2)
  \right]
\ ,
\label{div:Bmunu 2}
\end{eqnarray}
where $F_0$, $F_A$ and $F_3$ are defined by
\begin{eqnarray}
F_0 (s;M_1,M_2) &=& \int^1_0 dx \ln 
\left[ (1-x) M_1^2 + x M_2^2 - x (1-x) s \right] \ , \nonumber\\
F_A (s;M_1,M_2) &=& \int^1_0 dx\,(1-2x)\, \ln 
\left[ (1-x) M_1^2 + x M_2^2 - x (1-x) s \right] \ , \nonumber\\
F_3 (s;M_1,M_2) &=& \int^1_0 dx \, x (1-x) 
\ln \left[ (1-x) M_1^2 + x M_2^2 - x (1-x) s \right] 
\ .
\end{eqnarray}

We consider the quantum corrections denoted by $\Pi^{{\rm (vac)}\mu\nu}$
[see Eq.~(\ref{eq:TPart})].
At tree level
the two-point functions of $\overline{\cal A}_\mu$,
$\overline{\cal V}_\mu$ and $\overline{V}_\mu$ are given by
\begin{eqnarray}
 \Pi_\perp^{\rm{(tree)}\mu\nu}(p)
  &=& F_{\pi, \rm{bare}}^2 g^{\mu\nu} +
     2z_{2, \rm{bare}}(p^2 g^{\mu\nu} - p^\mu p^\nu)\ , \nonumber\\
 \Pi_\parallel^{\rm{(tree)}\mu\nu}(p)
  &=& F_{\sigma, \rm{bare}}^2 g^{\mu\nu} +
     2z_{1, \rm{bare}}(p^2 g^{\mu\nu} - p^\mu p^\nu)\ , \nonumber\\
 \Pi_V^{\rm{(tree)}\mu\nu}(p)
  &=& F_{\sigma, \rm{bare}}^2 g^{\mu\nu} -
     \frac{1}{g_{\rm{bare}}^2}
     (p^2 g^{\mu\nu} - p^\mu p^\nu)
\ ,\nonumber\\
 \Pi_{V\parallel}^{\rm{(tree)}\mu\nu}(p)
  &=& {}- F_{\sigma, \rm{bare}}^2 g^{\mu\nu} +
     z_{3, \rm{bare}}(p^2 g^{\mu\nu} - p^\mu p^\nu)
\ .
\label{TP tree}
\end{eqnarray}
Thus the one-loop contributions to
$\Pi_\perp^{{\rm (vac)}\mu\nu}$
give the quantum corrections to $F_\pi ^2$ and $z_2$.
Similarly,
each of the one-loop contributions to
$\Pi_\parallel^{{\rm (vac)}\mu\nu}$, $\Pi_V^{{\rm (vac)}\mu\nu}$ and
$\Pi_{V\parallel}^{{\rm (vac)}\mu\nu}$ includes the quantum corrections to 
two parameters shown above.
For distinguishing the quantum corrections to two parameters included
in the two-point function,
it is convenient to decompose each two-point function as
\begin{equation}
 \Pi^{{\rm (vac)}\mu\nu}(p)=\Pi^{{\rm (vac)}S}(p)g^{\mu\nu} +
                 \Pi^{{\rm (vac)}LT}(p)(g^{\mu\nu}p^2 - p^\mu p^\nu).
\label{decomp T0}
\end{equation}
It should be noticed that, since we use the Lagrangian with Lorentz
invariance, the form of the quantum corrections is Lorentz invariant.
Then, the relation between four components given in 
Eqs.~(\ref{Pi perp decomp})-(\ref{eq:VV-decompose}) and two components
shown above are given by
\begin{eqnarray}
 &&\Pi^{{\rm (vac)}t} = \Pi^{{\rm (vac)}s} = \Pi^{{\rm (vac)}S},
\nonumber\\
 &&\Pi^{{\rm (vac)}L} = \Pi^{{\rm (vac)}T} = \Pi^{{\rm (vac)}LT}.
\end{eqnarray}
Using the decomposition in Eq.~(\ref{decomp T0}), we identify 
$\Pi^{{\rm (vac)}S}_{\perp\mbox{\scriptsize(1-loop)}}(p^2)$ with the quantum
correction to $F_\pi^2$, 
$\Pi^{{\rm (vac)}LT}_{\perp\mbox{\scriptsize(1-loop)}}(p^2)$ 
with that to $z_2$,
and so on.
It should be noticed that
the following relation is satisfied~\cite{HYa,HYc}:
\begin{equation}
\Pi_V^{{\rm (vac)}S}(p^2) = \Pi_\parallel^{{\rm (vac)}S}(p^2) =
- \Pi_{V\parallel}^{{\rm (vac)}S}(p^2) \ .
\label{Pi V S equal}
\end{equation}
Then the quantum correction to $F_\sigma^2$ can be extracted from
any of $\Pi_\parallel^{\mu\nu}$,
$\Pi_V^{\mu\nu}$ and $\Pi_{V\parallel}^{\mu\nu}$.

We note that in Refs.~\cite{HYa,HYc} the finite corrections
of $\Pi^{{\rm (vac)}\mu\nu}_{\mbox{\scriptsize(1-loop)}}$ are neglected
by assuming that they are small.
In this paper
we include these finite contributions
in addition to the divergent corrections.
As in Refs.~\cite{HS,HSasaki},
we adopt the on-shell renormalization condition.
They are expressed as
\footnote{
  In the framework of the ChPT with HLS, 
  the renormalization point $\mu$ should be taken 
  as $\mu \geq M_\rho$
  since the vector meson decouples at $\mu = M_\rho$.
  Below the scale $M_\rho$ the parameter $F_\pi$, which is expressed 
  as $F_\pi^{(\pi)}(\mu)$ in Refs.~\cite{HYa,HYc}, runs by the 
  quantum correction from the pion loop 
  alone.
  Then
  $F_\pi^2(\mu=0)$ 
  in the right-hand-side of the renormalization condition 
  Eq.~(\ref{Fpi renorm cond}) is defined by
  $[F_\pi^{(\pi)}(\mu=0)]^2$.
  Due to the presence of the quadratic divergence $F_\pi(M_\rho)$
  is not connected smoothly to $F_\pi^{(\pi)}(M_\rho)$.
  The relation between them is expressed as~\cite{HYa,HYc}
  $ [F_\pi^{(\pi)}(M_\rho)]^2 = F_\pi^2(M_\rho) +
    [ N_f/(4\pi)^2] [ a(M_\rho)/2 ] M_\rho^2$,
  where $a(\mu)$ for $\mu>M_\rho$ is defined as
  $ a(\mu) \equiv F_\sigma^2(\mu)/ F_\pi^2(\mu)$.
  By using this relation, the renormalization condition
  (\ref{Fpi renorm cond}) determines the condition
  for $F_\pi^2(M_\rho)$.
  Strictly speaking,
  we should use $F_\pi(M_\rho)$ in the calculations in the present analysis.
  However, the difference between $F_\pi(0)$ and $F_\pi(M_\rho)$ 
  inside the loop correction coming from the finite renormalization
  effect is of higher order, and we use $F_\pi(0)$ 
  inside one-loop corrections in this paper.
}:
\begin{eqnarray}
 \mbox{Re}\Bigl[\Pi_\perp^{{\rm (vac)}S} (p^2=0)\Bigr]
  &=& F_\pi^2 (\mu = 0)\ , 
\label{Fpi renorm cond}
\\
 \mbox{Re}\Bigl[\Pi_V^{{\rm (vac)}S} (p^2=M_\rho^2)\Bigr]
  &=& F_\sigma^2 (\mu = M_\rho), 
\label{Fs ren cond}
\\
 \mbox{Re}\Bigl[\Pi_V^{{\rm (vac)}LT} (p^2=M_\rho^2)\Bigr]
  &=& - \frac{1}{g^2(\mu = M_\rho)},
\label{g ren cond}
\end{eqnarray}
where $\mu$ denotes the renormalization point.
Using the above renormalization conditions,
we can write the two-point functions as
\begin{eqnarray}
\Pi_\perp^{{\rm (vac)}S}(p^2) &=&
  F_\pi^2(0) + \widetilde{\Pi}_\perp^S(p^2) \ ,
\nonumber\\
\Pi_V^{{\rm (vac)}S}(p^2) &=& 
  F_\sigma^2(M_\rho) + \widetilde{\Pi}_V^S(p^2) \ ,
\nonumber\\
\Pi_V^{{\rm (vac)}LT}(p^2) &=& 
  - \frac{1}{g^2(M_\rho)} + \widetilde{\Pi}_V^{LT}(p^2) \ ,
\label{finite renormalization effects}
\end{eqnarray}
where $\widetilde{\Pi}_\perp^S(p^2)$,
$\widetilde{\Pi}_V^S(p^2)$ and $\widetilde{\Pi}_V^{LT}(p^2)$
denote the finite renormalization effects.
Their explicit forms are listed below.
{}From the above renormalization conditions
they satisfy
\begin{equation}
\widetilde{\Pi}_\perp^S(p^2=0)
=
\mbox{Re}\,\widetilde{\Pi}_V^S(p^2=M_\rho^2)
=
\mbox{Re}\,\widetilde{\Pi}_V^{LT}(p^2=M_\rho^2)
= 0
\ .
\label{zero FRE}
\end{equation}
Thus in the present renormalization scheme,
all the quantum effects for the on-shell parameters at leading order
are included through the renormalization group equations.

In the following, using the above functions, we summarize
the quantum corrections to two components
$\Pi^{{\rm (vac)}S}$ and $\Pi^{{\rm (vac)}LT}$ of the two-point functions
$\Pi_\perp^{{\rm (vac)}\mu\nu},\,\Pi_\parallel^{{\rm (vac)}\mu\nu},\,
\Pi_V^{{\rm (vac)}\mu\nu}$
and $\Pi_{V\parallel}^{{\rm (vac)}\mu\nu}$ which are defined in 
Eq.~(\ref{decomp T0}).
For $\overline{\cal A}_\mu$-$\overline{\cal A}_\nu$ two-point function,
we obtain
\begin{eqnarray}
 \Pi_{\perp\mbox{\scriptsize(1-loop)}}^{{\rm (vac)}S}(p^2)
  &=& - \frac{N_f}{(4\pi)^2}
    \Bigl[ \frac{2-a}{2}\Lambda^2 + \frac{3}{4}a\,M_\rho^2 \ln\Lambda^2 
     {}+ \frac{1}{4}a\,M_\rho^2 \ln M_\rho^2 \nonumber\\
  &&\qquad\qquad{}- a\,M_\rho^2 \Bigl\{
    1 + F_0(p^2;M_\rho,0) + \frac{1}{4}F_A(p^2;M_\rho,0)
    \Bigr\} \Bigr], \nonumber\\
 \Pi_{\perp\mbox{\scriptsize(1-loop)}}^{{\rm (vac)}LT}(p^2)
  &=& - \frac{N_f}{(4\pi)^2}\frac{a}{4}
    \Bigl[ \frac{1}{3}\ln\Lambda^2 - F_0(p^2;M_\rho,0) + 
      4F_3(p^2;M_\rho,0) \Bigr]. 
\end{eqnarray}
Corrections to
$\overline{\cal V}_\mu$-$\overline{\cal V}_\nu$ two-point function
are given by
\begin{eqnarray}
 \Pi_{\parallel\mbox{\scriptsize(1-loop)}}^{{\rm (vac)}S}(p^2)
  &=& -\frac{N_f}{(4\pi)^2}
   \Bigl[ \frac{a^2 + 1}{4}\Lambda^2 + \frac{3}{4}M_\rho^2\ln M_\rho^2
    {}+ M_\rho^2 \Bigl\{ \frac{1}{4}\ln M_\rho^2 - 1 -
    F_0(p^2;M_\rho,M_\rho) \Bigr\}\Bigr], \nonumber\\
 \Pi_{\parallel\mbox{\scriptsize(1-loop)}}^{{\rm (vac)}LT}(p^2)
  &=& -\frac{N_f}{(4\pi)^2}\frac{1}{8}
   \Bigl[ \frac{a^2 - 4a + 5}{3}\ln\Lambda^2 - F_0(p^2;M_\rho,M_\rho)
\nonumber\\ 
  &&\qquad\qquad\qquad\qquad{}+ 4F_3(p^2;M_\rho,M_\rho) - 
    4(2-a)^2 \ln M_\rho^2 \Bigr]. 
\end{eqnarray}
As for $\overline{V}_\mu$-$\overline{V}_\nu$ we have
\begin{eqnarray}
 \Pi_{V\mbox{\scriptsize(1-loop)}}^{{\rm (vac)}S}(p^2) 
  &=& \Pi_{\parallel\mbox{\scriptsize(1-loop)}}^{{\rm (vac)}S}(p^2), 
\nonumber\\
 \Pi_{V\mbox{\scriptsize(1-loop)}}^{{\rm (vac)}LT}(p^2)
  &=& -\frac{N_f}{(4\pi)^2}
   \Bigl[ \frac{a^2 - 87}{24}\ln\Lambda^2 + 4 + 
     \frac{23}{8}F_0(p^2;M_\rho,M_\rho) \nonumber\\ 
  &&\qquad\qquad{}+ 
     \frac{9}{2}F_3(p^2;M_\rho,M_\rho) -
     \frac{a^2}{8} \Bigl\{ F_0(p^2;0,0) - 4F_3(p^2;0,0) \Bigr\}\Bigr].
\end{eqnarray}
Finally, corrections to $\overline{V}_\mu$-$\overline{\cal V}_\nu$
two-point function are expressed as
\begin{eqnarray}
 \Pi_{V\parallel\mbox{\scriptsize(1-loop)}}^{{\rm (vac)}S}(p^2)
  &=& - \Pi_{\parallel\mbox{\scriptsize(1-loop)}}^{{\rm (vac)}S}(p^2), 
\nonumber\\
 \Pi_{V\parallel\mbox{\scriptsize(1-loop)}}^{{\rm (vac)}LT}(p^2)
  &=& -\frac{N_f}{(4\pi)^2}\frac{1}{8}
   \Bigl[ \frac{1 + 2a - a^2}{3} \ln\Lambda^2 - a(2-a)\ln M_\rho^2
\nonumber\\
  &&\qquad\qquad\qquad\qquad {}- F_0(p^2;M_\rho,M_\rho) +
   4F_3(p^2;M_\rho,M_\rho) \Bigr]. 
\end{eqnarray}

The renormalization conditions in Eqs.~(\ref{Fpi renorm cond})-
(\ref{g ren cond}) lead to the following relations among the bare and
renormalized parameters:
\begin{eqnarray}
 &&F_{\pi,\rm{bare}}^2-\frac{N_f}{4(4\pi)^2}
 \bigl[ 2(2-a)\Lambda^2 + 3aM_\rho^2 \ln\Lambda^2
 \bigr] \nonumber\\
 &&\qquad = F_\pi^2(0) - \frac{N_f}{4(4\pi)^2}aM_\rho^2
    \bigl[ 3\ln M_\rho^2 + \frac{1}{2}
    \bigr], \\
 &&F_{\sigma,\rm{bare}}^2 - \frac{N_f}{4(4\pi)^2}
  \bigl[ (a^2 + 1)\Lambda^2 + 3M_\rho^2 \ln\Lambda^2
  \bigr] \nonumber\\
 &&\qquad = F_\sigma^2(M_\rho) - \frac{N_f}{4(4\pi)^2}M_\rho^2
    \bigl[ 3\ln M_\rho^2 - 4(1-\sqrt{3}\tan^{-1}\sqrt{3})
    \bigr], \\
 &&\frac{1}{g_{\rm{bare}}^2} - \frac{N_f}{(4\pi)^2}
  \frac{87-a^2}{24}\ln\Lambda^2 \nonumber\\
 &&\qquad = \frac{1}{g^2(M_\rho)} - \frac{N_f}{8(4\pi)^2}
    \bigl[ \frac{87-a^2}{3}\ln M_\rho^2 - \frac{147-5a^2}{9}+
           41\sqrt{3}\tan^{-1}\sqrt{3}
    \bigr]. 
\end{eqnarray}

{}From these relations,
we obtain the RGEs for the parameters $F_\pi,\, g$ and $a$ 
as~\cite{HYa}
\begin{eqnarray}
 \mu \frac{d{F_\pi}^2}{d\mu} 
 &=& \frac{N_f}{2(4\pi)^2}
     \Bigl[ 3a^2 g^2 {F_\pi}^2 + 2(2-a){\mu}^2\Bigr],
 \label{eq:RGEF}\\
 \mu \frac{da}{d\mu} 
 &=& -\frac{N_f}{2(4\pi)^2}(a-1)
     \Bigl[ 3a(a+1)g^2 - (3a-1)\frac{{\mu}^2}{{F_\pi}^2} \Bigr],
 \label{eq:RGEa}\\
 \mu \frac{dg^2}{d\mu} 
 &=& -\frac{N_f}{2(4\pi)^2}\frac{87-a^2}{6}g^4.
 \label{eq:RGEg}
\end{eqnarray}

The finite renormalization effects of the two-point functions
are expressed as
\begin{eqnarray}
 \tilde{\Pi}_\perp^S(p^2)
 &=& \frac{N_f}{(4\pi)^2}a M_\rho^2 
    \Bigl[ -\Bigl( 1-\frac{M_\rho^2}{4p^2} \Bigr)
     \Bigl\{ 1-\Bigl( 1-\frac{M_\rho^2}{p^2} \Bigr)
      \ln \Bigl( 1-\frac{p^2}{M_\rho^2} \Bigr)\Bigr\} -
     \frac{1}{8} \Bigr] ,\\
 \tilde{\Pi}_V^S(p^2)
 &=& \frac{N_f}{(4\pi)^2}M_\rho^2
    \Bigl[ -\sqrt{3}\tan^{-1}\sqrt{3} +
     2\sqrt{\frac{4M_\rho^2 - p^2}{p^2}}
      \tan^{-1}\sqrt{\frac{p^2}{4M_\rho^2 - p^2}} \Bigr]\ ,
\label{C.2}
\\
 \tilde{\Pi}_V^{LT}(p^2)
 &=& \frac{N_f}{8(4\pi)^2}
    \Bigl[ \frac{a^2}{3}\ln \Bigl( \frac{-p^2}{M_\rho^2} \Bigr)
      {}- 24\Bigl( 1-\frac{M_\rho^2}{p^2} \Bigr) +
     41\sqrt{3}\tan^{-1}\sqrt{3} 
\nonumber \\
 &&\qquad\qquad 
    {}- \frac{2(12M_\rho^2 + 29p^2)}{p^2}
        \sqrt{\frac{4M_\rho^2 - p^2}{p^2}}
        \tan^{-1}\sqrt{\frac{p^2}{4M_\rho^2 - p^2}} \Bigr] .
\label{C.3}
\end{eqnarray}


\section{Functions}
\label{app:Functions}

In this appendix, we list the integral forms of the
functions which appear in 
the expressions of the physical quantities and the
several limits of the loop integrals shown in
Appendix~\ref{app:LIFT}.
The functions $\tilde{I}_{n}(T)$ and
$\tilde{J}^n_{m}(M ;T)$ ($n$, $m$: integers)
are given by
 \begin{eqnarray}
  &&\qquad\quad \tilde{I}_{n}(T) 
   = \int \frac{d^3 k}{(2\pi)^3}\frac{|\vec{k}|^{n-3}}{e^{k/T}-1}
   = \frac{1}{2\pi^2}\hat{I}_{n} T^n \ , \nonumber\\
  &&\qquad\qquad\qquad \hat{I}_{2} = \frac{{\pi}^2}{6},\quad
  \hat{I}_{4} = \frac{{\pi}^4}{15}\ , 
\label{I fun}
\\
  &&\quad \tilde{J}^n_{m}(M ;T) 
   = \int \frac{d^3 k}{(2\pi)^3} \frac{1}{e^{\omega /T}-1}
      \frac{|\vec{k}|^{n-2}}{{\omega}^m}\ , 
\label{J fun}
\end{eqnarray}
where
\begin{equation}
  \omega = \sqrt{k^2 + {M}^2}\ .
\end{equation}
The functions $\tilde{F}^n_{3}(p_0;M;T)$ and
$\tilde{G}_{n}(p_0;T)$, which appear in the vector meson pole
mass in section~\ref{sec:VMM}, are defined as
\begin{eqnarray}
  &&\tilde{F}^n_{3}(p_0;M;T) 
   = \int \frac{d^3 k}{(2\pi)^3}\frac{1}{e^{\omega /T}-1}
      \frac{4|\vec{k}|^{n-2}}{\omega (4{\omega}^2 - {p_0}^2)}\ , 
\nonumber\\
  &&\quad \tilde{G}_{n}(p_0;T) 
   = \int \frac{d^3 k}{(2\pi)^3}\frac{|\vec{k}|^{n-3}}{e^{k/T}-1}
       \frac{4|\vec{k}|^2}{4|\vec{k}|^2 - {p_0}^2}\ .
\end{eqnarray}


\end{document}